\documentclass[
 reprint,longbibliography,
showpacs,preprintnumbers,
bibnotes,
 amsmath,amssymb,
prb,
]{revtex4-2}

\usepackage{graphicx}
\usepackage[english]{babel}
\usepackage{microtype}       
\usepackage[utf8]{inputenc}
\usepackage[breaklinks=true,colorlinks=true,linkcolor=blue,urlcolor=blue,citecolor=blue]{hyperref}

\usepackage[dvipsnames]{xcolor}      
\usepackage{mathtools}
\mathtoolsset{showonlyrefs=true}
\usepackage{braket}
  \usepackage{booktabs}
\usepackage{tikz}
\usepackage{hhline}
\usepackage{bbm}

\usepackage{placeins}

\usetikzlibrary{shapes.geometric, arrows}
\usetikzlibrary{positioning, arrows.meta, calc, fit}
\tikzstyle{box} = [rectangle, rounded corners, 
minimum width=3cm, 
minimum height=1cm, 
text centered, 
text width=7cm, 
draw=black, 
fill=white]
\tikzstyle{arrow} = [thick,->,>=stealth]
\tikzstyle{darrow} = [thick,<->,>=stealth]

 \usepackage{longtable}

 \usepackage{siunitx}
 \usepackage{csquotes}
 
 \usepackage{multirow}
\usepackage{amssymb}

\usepackage{dcolumn}
\usepackage{bm}

\graphicspath{{figures/}}

\makeatletter
\g@addto@macro\bfseries{\boldmath}
\newcommand*{\balancecolsandclearpage}{%
   \close@column@grid
   \clearpage
   \twocolumngrid
 }
\makeatother

\begin{document}

\title{Real-time Dyson expansion for the nonequilibrium $GW$ approximation:\\ Correlated spectra and dependence on the reference propagator}

\author{Erik Schroedter}
\author{Jan-Philip Joost}
\author{Michael Bonitz
 \email{bonitz@theo-physik.uni-kiel.de}}
\affiliation{
Institute for Theoretical Physics and Astrophysics, Kiel University, 24098 Kiel, Germany \\ and Kiel Nano, Surface and Interface Science KiNSIS, Kiel University, Germany
}
\author{Vojtech Vlcek}\affiliation{Department of Chemistry and Biochemistry, University of California, Santa Barbara, California, USA}
\affiliation{Materials Department, University of California, Santa Barbara, California, USA}

\date{\today}

\begin{abstract}
Electronic spectra provide direct insight into the excitations and correlations of condensed matter systems and are the central observable of time- and angle-resolved photoemission experiments on driven quantum materials. Their description requires electron correlations beyond mean field. In equilibrium, the $GW$ approximation has become the method of choice for many materials, capturing quasiparticle renormalization and satellite structures through dynamical screening. Extending this approximation to nonequilibrium, however, is challenging. Full two-time $GW$ simulations within the nonequilibrium Green's functions approach scale at least cubically with propagation time, whereas improved scaling schemes such as time-local (adiabatic) approximations or the generalized Kadanoff--Baym ansatz (GKBA) and the G1--G2 scheme [Schluenzen \emph{et al.}, Phys. Rev. Lett. \textbf{124}, 076601 (2020)] usually retain only mean-field character in the spectra. The recently introduced real-time Dyson expansion (RT-DE) [Reeves and Vlcek, Phys. Rev. Lett. \textbf{133}, 226902 (2024)] recovers dynamical correlations in the spectrum at time-linear cost, but has so far been restricted to the second-order Born approximation with mean-field reference propagators. Here we formulate and implement the RT-DE for the nonequilibrium $GW$ self-energy with reference propagators of general form whose off-diagonal evolution is time local. In particular, we assess Hartree--Fock propagators, propagators with statically screened exchange with nonequilibrium screening, and a correlated propagator based on the Hartree--Fock GKBA (HF-GKBA). Benchmarks against exact diagonalization for a driven two-band lattice model with long-range interactions show that the mean-field and statically screened references yield the most accurate spectra, including satellite structures absent at mean-field level, whereas the GKBA-based approach best captures scattering-induced occupation dynamics, but artificially broadens and splits spectral peaks. When applied to large systems that are beyond the reach of exact methods, the scheme resolves the excitonic replica of the valence band and satellites identified as exciton shake-up, as well as their reshaping with increasing excitation density.
\end{abstract}
\maketitle

\section{Introduction}
The excitation of quantum materials by short laser pulses gives access to metal--insulator transitions, light-induced phase transitions and transient states that have no equilibrium counterpart~\cite{beebe2017, disa2023, bao2022, nuske2020,joost_prr_25}. The central experimental probe of these dynamics is time- and angle-resolved photoemission spectroscopy (trARPES), which follows band structures, gaps and satellite features on femtosecond time scales and images photoexcited exciton populations directly in momentum space~\cite{boschini_rmp_24, sie2019, man2021, dong2021}. The quantity measured in such experiments is the time-resolved spectral function \cite{Perfetto2016}. It shows the energies at which electrons can be removed from the evolving system, weighted by the transient occupations. Unlike observables such as densities or currents, this information is not contained in the instantaneous state of the system but in the correlations between processes at different times.

The natural theoretical framework for this is provided by nonequilibrium Green's functions (NEGF)~\cite{stefanucci_nonequilibrium_2013, balzer-book}. The formalism propagates the two-time Green's function, i.e., the quantity the spectral function is built from, via the (Keldysh--)Kadanoff--Baym equations (KBE), with correlations entering through a self-energy that can be improved systematically by diagrammatic expansion \cite{schluenzen_cpp16}. In particular, the NEGF framework extends the $GW$ approximation~\cite{hedin_new_1965,baym_conservation_1961}---the method of choice for the equilibrium spectra of many extended systems \cite{Martin_Reining_Ceperley_2016}, where the dynamically screened interaction generates the quasiparticle renormalization of the band structure and the satellite features arising from the coupling to charge-density fluctuations---to driven systems, in which screening becomes time dependent and follows the excited carrier distribution~\cite{perfetto2020}. Full two-time $GW$ simulations of this kind have been applied successfully to correlated systems out of equilibrium~\cite{von_friesen_successes_2009, Stan_2009,myohanen_2008,myohanen_prb_09,PuigvonFriesen2010}. These strengths, however, come at a substantial cost. The memory integrals in the collision term cause the runtime to grow at least cubically with the number of time steps $N_t$, and the two-time functions must be stored over the full propagation, which together restrict such simulations to short times and small systems. Moreover, full self-consistency is not automatically the more accurate choice as, for example, two-time KBE dynamics of small clusters exhibits artificial damping~\cite{von_friesen_successes_2009,PuigvonFriesen2010,schluenzen_prb17_comment}, and self-consistency has been shown not to guarantee improved time-resolved spectral functions~\cite{blommel2026}.

Considerable effort has therefore been devoted to the derivation of schemes that have an improved scaling with  $N_t$. For example, time-local (adiabatic) approximations discard the memory integrals entirely~\cite{attaccalite2011, attaccalite2013}. A different approach is given by the generalized Kadanoff--Baym ansatz (GKBA)~\cite{lipavsky_generalized_1986} where the off-diagonal Green's function is reconstructed from the time diagonal using mean-field propagators. This reduces the scaling to $\mathcal{O}(N_t^2)$ for the second-order Born approximation but yields no reduction for self-energies that resum entire diagram classes, such as $GW$. Time-linear scaling for these approximations is achieved by the G1--G2 scheme~\cite{schluenzen_prl_20,joost_prb_20, bonitz_pssb23}, which propagates the time-diagonal single- and two-particle Green's functions as coupled ordinary differential equations, and which has been extended to the $T$-matrix and dynamically screened ladder self-energies~\cite{joost_prb_22}. The $\delta$NEGF approach~\cite{schroedter_26} pushes the accessible system sizes beyond the limitations of the G1--G2 scheme by replacing the two-particle correlation function with an ensemble of mean-field-type trajectories, extending $GW$-level simulations to basis sizes of order $10^4$. However, so far it has only been formulated for the time-diagonal single-particle Green's function. The common limitation of the aforementioned approaches for spectroscopy is that the off-diagonal propagation remains of mean-field type. Dynamical quasiparticle renormalization and satellite structures are therefore absent from the resulting spectra. A different route to reducing the numerical scaling without significantly impacting the spectral accuracy is provided by compression approaches. Tensor-train representations of the two-time NEGF~\cite{Murray_2024, Sroda_2025} and low-rank approaches~\cite{Kaye_2021} substantially reduce the numerical costs of NEGF simulations, but part of the memory structure of the KBEs---and with it the unfavorable non-linear time scaling---remains.

In equilibrium, an analogous trade-off between cost and self-consistency is familiar from many-body perturbation theory. One-shot schemes evaluate the self-energy with a fixed reference Green's function instead of iterating the Dyson equation to self-consistency, as in the $G_0W_0$ approach. The results often improve compared to a fully self-consistent solution and yield high-quality spectra. However, they depend on the chosen starting point~\cite{van_setten_gw100_2015,golze_gw_2019,Korzdorfer2012,BrunevalMarques2013,ChenPasquarello2014}.

The recently introduced real-time Dyson expansion (RT-DE)~\cite{reeves2024} transfers the one-shot concept to the time domain. The self-energy is evaluated with a time-local reference propagator, which allows the memory integrals to be eliminated exactly in favor of coupled equations of motion for single- and two-particle functions, in close analogy to the G1--G2 scheme \cite{schluenzen_prl_20,joost_prb_20}. The resulting scheme scales linearly with the propagation time while retaining dynamical correlations in the off-diagonal direction, and thus in the spectra. Within the second-order Born approximation, the RT-DE has been shown to reproduce correlation-induced spectral features in good agreement with exact benchmarks, and to remain reliable over a broad range of couplings and excitation strengths~\cite{reeves2024, reeves2025, blommel2026}.

So far, however, the RT-DE has not been implemented and tested for self-energies with dynamical screening, although the corresponding $GW$ equations were derived for mean-field references in Ref.~\cite{reeves2024}. Given the role of screening for the spectra of extended systems, the $GW$ level is the natural target for a spectroscopy method. Moreover, all applications so far have only considered mean-field (Hartree--Fock) reference propagators. By analogy with the starting-point dependence of equilibrium one-shot schemes, the reference must be expected to influence the resulting spectra.

In this work we formulate the RT-DE equations of motion for the $GW$ self-energy with a general time-local reference propagator and present their first implementation and benchmarks. Beyond the mean-field reference, we consider two further propagators given by a statically screened (HSEX) propagator, whose screened interaction is rebuilt at each time step and thus carries nonequilibrium screening, and a propagator based on the HF-GKBA, which supplies a correlated time diagonal. Benchmarks against exact diagonalization for a driven two-band lattice model with long-range interactions, in and out of equilibrium and at two fillings, highlight the importance of the choice of reference. Applications to larger systems demonstrate that the approach resolves excitonic replicas and satellites identified as exciton shake-up at system sizes far beyond the reach of exact methods.

This paper is structured as follows. Section~\ref{s:NEGF} summarizes the NEGF framework, the $GW$ approximation and the time-resolved spectral function. Section~\ref{s:RTDE} introduces the RT-DE and derives the equations of motion for general time-local references. Section~\ref{s:reference_propagators} presents the three reference propagators. Section~\ref{s:numerics} contains the numerical results, and Sec.~\ref{s:conclusion} summarizes our conclusions and discusses future directions.

\section{Theoretical Framework} \label{s:NEGF}
We briefly summarize the nonequilibrium Green's function (NEGF) framework and introduce the notation used throughout this work. For a detailed overview we refer to Refs.~\cite{stefanucci_nonequilibrium_2013, balzer-book}.
 
We consider a system of fermions described by a generic Hamiltonian of the form
\begin{align}
    \hat{H}(t) = \sum_{ij}h_{ij}(t) \hat{c}_i^\dagger \hat{c}_j + \frac{1}{2}\sum_{ijkl}w_{ijkl}(t)\hat{c}^\dagger_i\hat{c}^\dagger_j\hat{c}_l\hat{c}_k,
\end{align}
containing a single-particle contribution $h$ and a pair interaction $w$. The operators $\hat{c}_i$ and $\hat{c}_i^\dagger$ annihilate and create a particle in the single-particle state $\psi_i$, where  $(\psi_i)_{i\in B}$ denotes an orthonormal basis of the underlying single-particle Hilbert space. The index $i$ is understood as a multi-index that describes all single-particle quantum numbers, for example, for the lattice model of Sec.~\ref{s:numerics}, it comprises the site, band and spin index.  Both contributions, $h(t)$ and $w(t)$, are allowed to be time dependent, which serves to describe the external excitation and, through adiabatic switching of $w$, to prepare a correlated initial state from an uncorrelated one~\cite{hermanns_hubbard_2014, schluenzen_cpp16}.
\subsection{Nonequilibrium Green's functions}
\begin{figure}[t]
  \centering
  \includegraphics[width=\columnwidth]{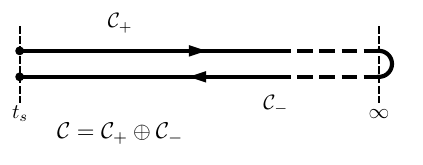}
  \caption{Keldysh time contour $\mathcal{C}=\mathcal{C}_+\oplus\mathcal{C}_-$
  used for the real-time propagation. The system is prepared at the switch-on
  time $t_s$. The upper (forward) branch $\mathcal{C}_+$ runs from $t_s$ toward
  $t\to\infty$ and the lower (backward) branch $\mathcal{C}_-$ returns.
  The two-time Green's function $G(t,t')$ is evaluated with both arguments on this
  contour.}
  \label{fig:contour}
\end{figure}

The single-particle nonequilibrium Green's function is defined as
\begin{align}
    G_{ij}(z,z') =  \frac{1}{\mathrm{i}\hbar}\big\langle \mathcal{T}_\mathcal{C}\big\{\hat{c}_i(z) \hat{c}^\dagger_j(z')\big\}\big\rangle,
\end{align}
where $\mathcal{T}_\mathcal{C}$ denotes the time-ordering operator on the Keldysh contour \cite{keldysh_diagram_1964,bonitz_pss_19_keldysh} $\mathcal{C}=\mathcal{C}_+\oplus\mathcal{C}_-$, consisting of a forward branch $\mathcal{C}_+=[t_s,\infty)$ and a backward branch $\mathcal{C}_-=-\,\mathcal{C}_+$, and $z,z'\in\mathcal{C}$ are contour times, cf. Fig.~\ref{fig:contour}.  The expectation value is taken with respect to the density operator of the initial state at time $t_s$. Real-time functions are recovered in the standard way: the lesser (greater) component $G^<$ ($G^>$) corresponds to real times $t$ on the forward (backward) and $t'$ on the backward (forward) branch,
\begin{align}
    G^<(t,t') &= G(t\in\mathcal{C}_+,t'\in\mathcal{C}_-),\\
    G^>(t,t') &= G(t\in\mathcal{C}_-,t'\in\mathcal{C}_+).
\end{align}
Further, the retarded and advanced components are given by
\begin{align}
    G^\mathrm{R}(t,t') &= \Theta(t-t')\{G^>(t,t')-G^<(t,t')\},\\
    G^\mathrm{A}(t,t') &= \Theta(t'-t)\{G^<(t,t')-G^>(t,t')\}.
\end{align}
The time evolution of the single-particle NEGF is described by the (Keldysh--)Kadanoff--Baym equation (KBE),
\begin{align}
    \mathrm{i}\hbar\partial_z G(z,z')=\delta_\mathcal{C}(z,z')+h(z)G(z,z')+I(z,z'),
\label{eq:g1-eq}
\end{align}
where the collision term,
\begin{align}
    I_{ij}(z,z') = -\mathrm{i}\hbar\sum_{pqr}w_{ipqr}(z)G^{(2)}_{rqpj}(z,z'),
    \label{eq:collision_G2}
\end{align}
couples $G$ to the two-particle NEGF,
\begin{align}
    G^{(2)}_{ijkl}(z,z') &= G^{(2)}_{ijkl}(z,z,z^+,z'),\\
    G^{(2)}_{ijkl}(z_1,z_2,z'_1,z'_2)&= -\frac{1}{\hbar^2}\left\langle \mathcal{T}_\mathcal{C}\left\{\hat{c}_i(z_1)\hat{c}_j(z_2)\hat{c}^\dagger_l(z'_2)\hat{c}^\dagger_k(z'_1)\right\}\right\rangle,
\end{align}
where $z^+=z+\epsilon$ denotes a time infinitesimally later on the contour, which fixes the equal-time operator ordering.
Since the equation of motion for $G^{(2)}$ contains, in turn, the three-particle Green's function $G^{(3)}$, one obtains the Martin--Schwinger hierarchy~\cite{martin_theory_1959}, which must be closed by approximations, either through a self-energy $\Sigma[G]$, cf.~Sec.~\ref{ss:selfenergy}, or by approximating the two-particle NEGF directly.
 
For the latter approach, it is useful to decompose the two-particle NEGF into mean-field and correlation contributions,
\begin{align}
    G^{(2)}=G^{(2),\mathrm{HF}}+\mathcal{G},
    \label{eq:G2_decomp}
\end{align}
where the Hartree--Fock part is given by
\begin{align}
    G^{(2),\mathrm{HF}}_{ijkl}(z_1,z_2,z'_1,z'_2)=&\, G_{ik}(z_1,z'_1)G_{jl}(z_2,z'_2)\\ &-G_{il}(z_1,z'_2)G_{jk}(z_2,z'_1).
\end{align}
Neglecting the correlation contribution, $\mathcal{G}\to 0$, yields the mean-field (Hartree--Fock) approximation. Equation~\eqref{eq:G2_decomp} is the starting point for the RT-DE formulation of Sec.~\ref{s:RTDE}.
 
\subsection{Self-energy and correlation contributions} \label{ss:selfenergy}

A formal closure of the hierarchy is achieved by introducing the self-energy $\Sigma=\Sigma[G]$, in terms of which the collision term \eqref{eq:collision_G2} is rewritten as
\begin{align}
    I(z,z')&=\int_\mathcal{C}\Sigma(z,\bar{z})G(\bar{z},z')\,\mathrm{d}\bar{z}\\&\equiv \big(\Sigma \cdot G\big)(z,z'),
\label{eq:col-int-sigma}
\end{align}
where a dot denotes a product that, in addition, includes a convolution along the Keldysh contour.

Analogously to the decomposition of $G^{(2)}$, we split the self-energy into a Hartree--Fock part and a correlation part,
\begin{align}
    \Sigma=\Sigma^\mathrm{HF}+\Sigma^\mathrm{cor}.
\end{align}
The Hartree--Fock self-energy,
\begin{align}
    \Sigma^\mathrm{HF}_{ij}(z,z')=-\mathrm{i}\hbar\delta_\mathcal{C}(z,z')\sum_{pq}w^-_{ipjq}(z)G_{qp}(z,z^+),
\end{align}
with the anti-symmetrized pair potential defined as $w^-_{ijkl}= w_{ijkl}-w_{ijlk}$, is singular on the contour, and can be absorbed into an effective single-particle Hamiltonian,
\begin{align}
    h^\mathrm{HF}= h+\Sigma^\mathrm{HF}.
\end{align}
The remaining, correlation part of the collision term then reads
\begin{align}
    \mathcal{I}_{ij}(z,z')&= \sum_p\int_\mathcal{C}\Sigma^\mathrm{cor}_{ip}(z,\bar{z})G_{pj}(\bar{z},z')\,\mathrm{d}\bar{z}\\
    &=-\mathrm{i}\hbar\sum_{pqr}w_{ipqr}(z)\mathcal{G}_{rqpj}(z,z'),
    \label{eq:cor_collision_G2}
\end{align}
and the equation for $G$, Eq.~\eqref{eq:g1-eq}, becomes
\begin{align}
    \mathrm{i}\hbar\partial_z G(z,z')=\delta_\mathcal{C}(z,z')+h^\mathrm{HF}(z)G(z,z')+\mathcal{I}(z,z'). \quad
    \label{eq:KBE_HF}
\end{align}
Expressing the collision term~\eqref{eq:cor_collision_G2} through the correlation self-energy and through the correlated part of the two-particle NEGF connects the two closures of the Martin--Schwinger hierarchy. Any diagrammatic approximation for $\Sigma^\mathrm{cor}$ induces a corresponding $\mathcal{G}$, and vice versa. The correlated part entering the RT-DE below is thus not an independent ansatz, but the one implied by the chosen self-energy.

The second form removes the memory integral from the equation of motion of the single-particle NEGF, Eq.~\eqref{eq:KBE_HF}, whose collision term is now a simple contraction of $w$ with $\mathcal{G}$. The memory, however, is not eliminated but shifted, and reappears in the equation for $\mathcal{G}$. Removing it there requires further approximations beyond the considered self-energy approximation, for example, time locality of the propagators entering these contour integrals at the two-particle level or the Markov limit \cite{bonitz_qkt,bonitz_cpp18}.
 
Self-energy approximations may be constructed as an expansion in the pair interaction $w$---first order yields the Hartree--Fock self-energy, second order the second-order Born approximation (SOA)---or by summing selected classes of diagrams to infinite order~\cite{kadanoff_quantum_1962, stefanucci_nonequilibrium_2013,schluenzen_jpcm_19}. We focus on the most prominent approximation of the latter type, the $GW$ approximation, although the following considerations generalize to other self-energy approximations.
 
\subsection{$GW$ approximation} \label{ss:GW}
The charge of an electron polarizes the surrounding system, which responds with a density rearrangement that partially compensates the original field. The interaction between two quasiparticles is therefore screened, leading to an effective interaction, whose range and strength differ substantially from those of the bare Coulomb repulsion. The $GW$ approximation takes this as its starting point and constructs the self-energy from the dynamically screened interaction $W$ rather than from $w$. It is the standard approximation for quasiparticle energies and photoemission spectra of extended systems~\cite{hedin_new_1965, golze_gw_2019}.

Within the $GW$ approximation, the exchange and correlation parts of the self-energy are combined into the exchange-correlation self-energy and approximated using the dynamically screened interaction $W$ as
\begin{align}
    \Sigma^\mathrm{xc}_{ij}(z,z')=\mathrm{i}\hbar\sum_{pq}W_{ipqj}(z,z')G_{qp}(z,z').
    \label{eq:sigma_xc}
\end{align}
Here, the screened interaction is given by
\begin{align}
    W(z,z')=w(z)\delta_\mathcal{C}(z,z')+w(z)\chi(z,z')w(z').
    \label{eq:W}
\end{align}
The reducible polarizability $\chi$ obeys the Bethe--Salpeter equation (BSE) given by
\begin{align}
    \chi=\chi^0+\chi^0\cdot w\cdot\chi,
    \label{eq:BSE_chi}
\end{align}
with the irreducible polarizability %
\begin{align}
    \chi^0_{ijkl}(z,z')= -\mathrm{i}\hbar\, G_{il}(z,z')G_{jk}(z',z). \label{eq:def_chi0}
\end{align}
Here and in the following, products of two-particle quantities $A_{ijkl},B_{ijkl}$ denote the contraction $(AB)_{ijkl}=\sum_{pq}A_{ipkq}B_{qjpl}$.

The BSE~\eqref{eq:BSE_chi} and the definition of $\chi^0$ \eqref{eq:def_chi0} are the formal realization of the aforementioned picture of dynamical screening. The irreducible polarizability $\chi^0$ describes the creation and propagation of a single particle--hole pair, and the BSE reiterates these ``polarization bubbles'' into the charge-density fluctuation induced in the surrounding electron system. 

The static part of the screened interaction renormalizes the effective single-particle structure, whereas its frequency dependence generates the dynamical correlation effects relevant for spectroscopy. In particular, coupling to charge-density excitations transfers spectral weight from quasiparticle peaks into satellites. In finite systems these appear as additional discrete peaks, while out of equilibrium their position and weight evolve with the time-dependent screening. Such structures cannot be generated by a purely time-local single-particle potential.

Note that the singular first term of $W$ in Eq.~\eqref{eq:W} reproduces the Fock (exchange) self-energy. The correlation self-energy entering Eq.~\eqref{eq:KBE_HF} is therefore given by the dynamical part alone,
\begin{align}
    \Sigma^\mathrm{cor}_{ij}(z,z')=\mathrm{i}\hbar\sum_{pq}\big[w\chi w\big]_{ipqj}(z,z')\,G_{qp}(z,z').
    \label{eq:sigma_cor_GW}
\end{align}
Inserting Eq.~\eqref{eq:sigma_cor_GW} into Eq.~\eqref{eq:cor_collision_G2} and comparing coefficients, one obtains for the correlated part of the two-particle Green's function,
\begin{align}
    &\mathcal{G}_{ijkl}(z,z') =  \label{eq:G2_GW_general}
    \\&-\mathrm{i}\hbar\sum_{pqrs}\int_\mathcal{C}\chi_{ipkq}(z,\bar{z})w_{qrps}(\bar{z})G_{jr}(z,\bar{z})G_{sl}(\bar{z},z')\,\mathrm{d}\bar{z}.
\end{align}

Although the NEGF formalism allows for systematic improvements beyond the mean-field level, correlated self-energy approximations introduce memory effects through the contour convolutions in Eqs.~\eqref{eq:cor_collision_G2} and \eqref{eq:BSE_chi}. As a consequence, the runtime of a full solution of the KBEs scales at least cubically with the number of time steps, which limits fully correlated NEGF simulations to short propagation times and small systems. The real-time Dyson expansion developed in the next section is designed to retain dynamical correlation effects in the spectral information while avoiding this unfavorable time scaling.
\subsection{Time-resolved spectral function} \label{ss:spectral_function}
The central observable of this work is the time-resolved spectral function, which is directly related to the photoemission signal measured in time-resolved pump--probe experiments. Its removal (occupied) and addition (unoccupied) parts are obtained from the lesser and greater components of the single-particle NEGF as \cite{eckstein_measuring_2008, freericks_09}
\begin{align}
    A^<_\delta(\omega,t_p) &= -\mathrm{i}\hbar\iint \mathrm{e}^{-\mathrm{i}\omega(t-t')}\, \mathcal{S}_{t_p,\delta} (t)\, \mathcal{S}_{t_p,\delta}(t')\notag\\&\hspace{7em}\times\mathrm{Tr}\big[G^<(t,t')\big]\,\mathrm{d}t\,\mathrm{d}t', \label{eq:A_lesser}\\
    A^>_\delta(\omega,t_p) &= \mathrm{i}\hbar\iint \mathrm{e}^{-\mathrm{i}\omega(t-t')}\, \mathcal{S}_{t_p,\delta} (t)\, \mathcal{S}_{t_p,\delta}(t')\notag\\&\hspace{7em}\times\mathrm{Tr}\big[G^>(t,t')\big]\,\mathrm{d}t\,\mathrm{d}t', \label{eq:A_greater}
\end{align}
where $\mathcal{S}_{t_p,\delta}$ is the probe pulse used in time-resolved photoemission experiments, centered at the probe time $t_p$ and with width $\delta$. Here, we assume the pulse to be a Gaussian distribution,
\begin{align}
    \mathcal{S}_{t_p,\delta}(t)= \frac{1}{\sqrt{2 \pi \delta^2}}\, \mathrm{e}^{-\frac{(t-t_p)^2}{2\delta^2}},
\end{align}
which is normalized to one.

The full spectral function is given by the sum of both contributions,
\begin{align}
    A_\delta (\omega, t_p) = A^>_\delta(\omega,t_p) + A^<_\delta(\omega,t_p).
\end{align}
The occupied part, $A^<_\delta$, weights each single-particle excitation with its occupation and, thus, gives the intensity measured in time-resolved photoemission, whereas the unoccupied part $A^>_\delta$ resolves the unoccupied states and corresponds to inverse photoemission. Their sum, $A_\delta$, acts as the single-particle density of states at the probe time and contains the correlation-induced renormalization of the levels, the broadening of the quasiparticle peaks, and the incoherent satellite weight. For periodic systems, a momentum-resolved spectral function $A_\delta(k,\omega,t_p)$ is obtained analogously, by replacing the trace in Eqs.~\eqref{eq:A_lesser} and \eqref{eq:A_greater} with the diagonal matrix element $G^\gtrless_k(t,t')$ in the momentum basis. This is the quantity measured in time- and angle-resolved photoemission (trARPES).

The probe duration $\delta$ determines the resolution of $A_\delta(\omega,t_p)$, meaning that a longer probe sharpens the spectral features but averages over a longer time interval, so that energy and time resolution cannot be improved simultaneously.
 
The spectral function is determined by the time off-diagonal structure of $G^\gtrless(t,t')$, in contrast to time-diagonal observables such as densities and occupations. An accurate description of correlation effects in the spectrum, therefore, requires a correlated treatment of the off-diagonal propagation. Further, due to the Gaussian probe envelope, only times within a window of width $T_\delta \sim \delta $ around the probe time $t_p$ contribute to the integrals. Consequently, the NEGF has to be computed only for the $N_{t,\delta}$ time steps inside the probe window, where, in general, $N_{t,\delta}\ll N_t$ is independent of the total propagation time, cf. Fig.~\ref{fig:rtde_schematic}, for illustration. 

The (exact) spectral functions are positive semi-definite, i.e., $A^\gtrless_\delta (\omega,t_p) \geq 0$, for all frequencies $\omega$ and probe times $t_p$. This property, however, is not guaranteed to be fulfilled for general self-energy approximations, and diagrammatic constructions that satisfy this constraint are discussed in Refs.~\cite{Stefanucci2014PSD, Uimonen2015PSD}. As shown there, one such positive approximation is given by $GW$.

\section{Real-Time Dyson Expansion} \label{s:RTDE}
 
\subsection{Time-local schemes and one-shot corrections}
\label{ss:motivation}
 
The unfavorable scaling of the full KBEs originates from the memory integrals in the collision term. Two established strategies have been developed to remove them. The first replaces the memory-dependent collision integral by a time-local (adiabatic) approximation, evaluating the self-energy with the time-diagonal Green's function alone \cite{attaccalite2011, attaccalite2013}. While this restores a favorable scaling, it does so by discarding retardation effects, so that the resulting dynamics are effectively of mean-field character, and dynamical correlations are absent by construction rather than merely underrepresented. The second strategy is the generalized Kadanoff--Baym ansatz (GKBA)~\cite{lipavsky_generalized_1986}, which reconstructs the off-diagonal NEGF from the time-diagonal dynamics. The most common choice is to use mean-field propagators (HF-GKBA), see, for example,~\cite{hermanns_hubbard_2014} or \cite{bonitz_pssb23} for an overview. The HF-GKBA allows for a time-local reformulation in terms of the two-particle Green's function---the so-called G1--G2 scheme \cite{schluenzen_prl_20,joost_prb_20}---which achieves linear scaling with the number of time steps, even for advanced self-energies such as $GW$ (see Sec.~\ref{ss:HFGKBA} for more details). In both cases, the off-diagonal propagation remains of mean-field type, so dynamical spectral features are not retained.  Overcoming this deficiency without reintroducing the cubic time scaling is the purpose of the present work.
 
In equilibrium, the problematic cubic scaling with the number of time steps for approximations beyond a mean-field description can be circumvented by considering an equivalent reformulation of the KBEs. Using the ideal Green's function $G^0$, the KBEs can be transformed into the Dyson equation,
\begin{align}
    G = G^0 + G^0 \cdot\Sigma[G] \cdot G.
    \label{eq:dyson}
\end{align}
In equilibrium, all two-time functions depend only on the time difference, so the contour convolutions become ordinary convolutions. Then, the Fourier transform can be used to solve the Dyson equation in frequency space, where the convolutions become ordinary products. This reduces the computational complexity significantly so that even large systems can be considered.

\textit{One-shot methods} are an alternative where the self-consistency of the Dyson equation is reduced by evaluating the self-energy only once, using $G^0$, instead of the full Green's function,
\begin{align}
    G \approx G^0 + G^0 \cdot\Sigma[G^0] \cdot G.
    \label{eq:dyson_oneshot}
\end{align}
Equivalently to the Dyson equation \eqref{eq:dyson} using the ideal Green's function $G^0$, it is possible to consider the mean-field (Hartree--Fock) solution $G^\mathrm{HF}$ as the starting point. This requires the full self-energy $\Sigma$ to be replaced by its correlation part $\Sigma^\mathrm{cor}$, since the Hartree--Fock contribution is then already contained:
\begin{align}
    G = G^\mathrm{HF}+G^\mathrm{HF}\cdot\Sigma^\mathrm{cor}[G]\cdot G.\label{eq:dyson_HF}
\end{align}
Again, one-shot corrections can be considered in this equivalent framework and $G^\mathrm{HF}$ is used as a one-shot reference instead:
\begin{align}
    G \approx G^\mathrm{HF}+G^\mathrm{HF}\cdot\Sigma^\mathrm{cor}[G^\mathrm{HF}]\cdot G.\label{eq:dyson_HF_oneshot}
\end{align}
Although both versions of the Dyson equation, Eqs.~\eqref{eq:dyson} and \eqref{eq:dyson_HF}, are fully equivalent, the two aforementioned one-shot equations \eqref{eq:dyson_oneshot} and \eqref{eq:dyson_HF_oneshot} are not equivalent, and the results differ depending on the reference \cite{van_setten_gw100_2015, golze_gw_2019,Korzdorfer2012,BrunevalMarques2013}.

One-shot approaches not only reduce the numerical cost, as the self-energy does not have to be reevaluated for every iteration, but have also been found to improve the quality of the spectral information in many cases. Unlike the fully self-consistent solution of the Dyson equation, however, one-shot approaches depend on the chosen reference. The real-time Dyson expansion (RT-DE), which was recently introduced in Ref.~\cite{reeves2024}, extends this one-shot concept to the nonequilibrium domain. The nonequilibrium RT-DE should therefore inherit a starting-point dependence, which we quantify below.
 
\subsection{Definition of the real-time Dyson expansion (RT-DE)}
\label{ss:RTDE_definition}
The RT-DE combines the one-shot idea with the two-particle Green's function representation of correlations introduced in Sec.~\ref{s:NEGF}. Its defining approximation is to evaluate the correlation self-energy in the
collision term \eqref{eq:cor_collision_G2} with a time-local reference Green's
function $G^\mathrm{loc}$:
\begin{align}
    \mathcal{I}_{ij}(z,z') &\approx \sum_p \int_\mathcal{C} \Sigma^\mathrm{cor}_{ip}[G^\mathrm{loc}](z,\bar z)\, G_{pj}(\bar z, z')\,
    \mathrm{d}\bar z \label{eq:RTDE_collision}\\
    &\equiv - \mathrm{i}\hbar \sum_{pqr} w_{ipqr}(z)\mathcal{G}_{rqpj}[G^\mathrm{loc}, G](z,z'), \label{eq:RTDE_G2_def}
\end{align}
and is thus the real-time analog of the one-shot Dyson equation~\eqref{eq:dyson_HF_oneshot}. Additionally, the same replacement is done for the singular part of the self-energy, i.e., $\Sigma^\mathrm{HF}[G]\rightarrow \Sigma^\mathrm{HF}[G^\mathrm{loc}]$.

The propagator used for the one-shot correction obeys a time-local equation of motion,
\begin{align}
    \mathrm{i}\hbar \partial_z G^\mathrm{loc}(z,z') =
    \delta_\mathcal{C}(z,z')+ h^\mathrm{loc}(z)\, G^\mathrm{loc}(z,z'),
    \label{eq:EOM_Gloc}
\end{align}
for some effective single-particle Hamiltonian $h^\mathrm{loc}(z)$ to be specified below. 
 
The replacement $\Sigma^\mathrm{cor}[G]\rightarrow\Sigma^\mathrm{cor}[G^\mathrm{loc}]$ does not by itself remove the memory integrals, as the collision term in Eq.~\eqref{eq:RTDE_collision} still contains a convolution over the full history. However, since all Green's functions entering the self-energy obey time-local equations of motion, the memory dependence of the integrand can be transferred to the two-particle function $\mathcal{G}^\gtrless(t,t')$ itself. The integral equation can then be reformulated as a coupled system of ordinary differential equations, which are propagated with respect to $t$ for fixed $t'$, analogously to the G1--G2 reformulation of the HF-GKBA \cite{schluenzen_prl_20, joost_prb_20}.

So far, the RT-DE has been formulated and applied within the second-order Born approximation~\cite{reeves2024, reeves2025}. In this case, it was shown to reproduce spectral features that are absent in mean-field descriptions, such as emergent excitonic peaks and satellite structures, in good agreement with exact results for small systems. In all of these applications, however, the reference propagator was restricted to the mean-field (Hartree--Fock) level. Systematic benchmarks against fully self-consistent KBE calculations and the HF-GKBA indicate that mean-field reference trajectories typically provide a reliable basis for the RT-DE over a broad range of interaction strengths and excitation densities~\cite{blommel2026}.  There are, however, situations in which a mean-field description of the time diagonal is insufficient. This is the case, for example, in homogeneous systems. Consider a spatially uniform field that couples only states of equal momentum, then single-particle states the field does not address directly can be populated only through scattering, which is absent without a collision term. Correlated time-diagonal schemes such as the HF-GKBA are not subject to these limitations, and the differences to the mean-field description can be substantial---especially for strong correlations or driving. So far, the effect of the reference propagator on the resulting spectra has not been investigated systematically. In this work we show that a correlated time diagonal does not by itself translate into improved spectra.

Furthermore, the RT-DE has not been applied to self-energy approximations involving dynamical screening. The corresponding  $GW$ equations were already derived for mean-field propagators in Ref.~\cite{reeves2024}. In the following, we give the expressions for a general time-local reference propagator, which follow from a slight modification of the original equations and their derivation, and allow all reference propagators considered in this work to be treated on the same footing.
 
\subsection{RT-DE equations for the $GW$ self-energy}
\label{ss:RTDE_GW}
 
Within the $GW$ approximation, the correlated part of the RT-DE two-particle Green's function follows from Eq.~\eqref{eq:G2_GW_general} as
\begin{align}
    &\mathcal{G}_{ijkl}(z,z') \equiv \mathcal{G}_{ijkl}[G^\mathrm{loc}, G](z,z')
    \\&= - \mathrm{i}\hbar\sum_{pqrs} \int_\mathcal{C} \chi_{ipkq}[G^\mathrm{loc}](z,\bar z)\, w_{qrps}(\bar z)\,G^\mathrm{loc}_{jr}(z,\bar z)\,\\&\hspace{11em} \times G_{sl}(\bar z, z')\,\mathrm{d}\bar z, \label{eq:G2_GW}
\end{align}
where the polarizability obeys the BSE with time-local propagators,
\begin{align}
 \chi = \chi^0[G^\mathrm{loc}]
 +\chi^0[G^\mathrm{loc}]\cdot w\cdot \chi.
 \label{eq:BSE_chi_loc}
\end{align}
Since Eq.~\eqref{eq:BSE_chi_loc} depends only on the time-local Green's function $G^\mathrm{loc}$, it can be transformed into a time-local equation of motion. The real-time components of $\chi$ obey
\begin{align}
    \mathrm{i}\hbar \partial_t \chi^\gtrless(t,t') =
    h^{(2),GW}(t)\, \chi^\gtrless(t,t'),
    \label{eq:EOM_chi}
\end{align}
with the effective two-particle Hamiltonian
\begin{align}
    h^{(2),GW}(t) &=
    h^{(2),\mathrm{loc}}(t)+h^{(2),\mathrm{cor}}(t).
    \label{eq:h2_GW}
\end{align}
Here we introduced the effective two-particle local Hamiltonian
\begin{align}
    h^{(2),\mathrm{loc}}_{ijkl}(t) &=
    h_{il}^\mathrm{loc}(t)\delta_{kj}-h^\mathrm{loc}_{jk}(t)\delta_{il},
\end{align}
which follows from the derivative of $\chi^0[G^\mathrm{loc}]$. Furthermore, arising from the derivative of the integral boundary, the effective Hamiltonian contains a correlation part describing polarization effects that is given by
\begin{align}
     h^{(2),\mathrm{cor}}(t) &=- \mathrm{i}\hbar\{\chi^{0,>}[G^\mathrm{loc}](t)- \chi^{0,<}[G^\mathrm{loc}](t) \}\, w(t).
\end{align}
Combining Eqs.~\eqref{eq:RTDE_collision}--\eqref{eq:EOM_chi}, the RT-DE for the $GW$ approximation transforms into a closed system of coupled equations of motion: the off-diagonal propagation of the single-particle and two-particle NEGF \cite{reeves2024},
\begin{widetext}
\begin{gather}
  \mathrm{i}\hbar \partial_t G^\gtrless_{ij}(t,t') = \sum_p h^\mathrm{HF}_{ip}(t)\,G_{pj}^\gtrless(t,t') - \mathrm{i}\hbar \sum_{pqr}w_{ipqr}(t)\, \mathcal{G}^\gtrless_{rqpj}(t,t'),    \label{eq:EOM_G1_RTDE-MF}\\
    \mathrm{i}\hbar \partial_t \mathcal{G}^\gtrless_{ijkl}(t,t') =
    \sum_{p}\big\{h^\mathrm{loc}_{ip}(t)\, \mathcal{G}^\gtrless_{pjkl}(t,t')
    + h^\mathrm{loc}_{jp}(t)\, \mathcal{G}^\gtrless_{ipkl}(t,t')
    -  \mathcal{G}^\gtrless_{ijpl}(t,t')\,h^\mathrm{loc}_{pk}(t)\big\}+ \Psi^\gtrless_{ijkl}(t,t') + \Pi^\gtrless_{ijkl}(t,t')\,,
    \label{eq:G2_GW_MF_propagators_1}
\end{gather}
together with the time-diagonal equations for $G^{\mathrm{loc},\gtrless}(t)$ and $\mathcal{G}^\mathrm{loc}(t)$,
\begin{gather}
    \mathrm{i}\hbar \partial_t G^{\mathrm{loc},\gtrless}_{ij}(t) =
    \sum_{p}\big\{ h^\mathrm{loc}_{ip}(t)\, G^{\mathrm{loc},\gtrless}_{pj}(t)
    - G^{\mathrm{loc},\gtrless}_{ip}(t)\,h^\mathrm{loc}_{pj}(t) \big\},
    \label{eq:EOM_Gloc_diag}\\
    \mathrm{i}\hbar \partial_t \mathcal{G}^\mathrm{loc}_{ijkl}(t) =
    \sum_{p}\big\{ h^\mathrm{loc}_{ip}(t)\,\mathcal{G}^\mathrm{loc}_{pjkl}(t)
    + h^\mathrm{loc}_{jp}(t)\, \mathcal{G}^\mathrm{loc}_{ipkl}(t)
    - \mathcal{G}^\mathrm{loc}_{ijpl}(t)\,h^{\mathrm{loc}}_{pk}(t)
    -\mathcal{G}^\mathrm{loc}_{ijkp}(t)\, h^\mathrm{loc}_{pl}(t)\big\}+\Psi^\mathrm{loc}_{ijkl}(t) + \Pi^\mathrm{loc}_{ijkl}(t).
    \label{eq:G2_GW_MF_propagators_2}
\end{gather}
Here, $\Psi$ denotes the second-order scattering term and $\Pi$ the polarization contribution:
\begin{gather}
    \Psi^\gtrless_{ijkl}(t,t') = -\hbar^2 \sum_{pqrs} \big\{G^{\mathrm{loc},>}_{ip}(t)G^{\mathrm{loc},>}_{jq}(t)G^{\mathrm{loc},<}_{rk}(t) - (>\leftrightarrow <) \big\}\, w_{pqrs}(t)\, G^\gtrless_{sl}(t,t'),
    \label{eq:def_Psi}\\
    \Psi^\mathrm{loc}_{ijkl}(t) = -\hbar^2 \sum_{pqrs}\big\{ G^{\mathrm{loc},>}_{ip}(t)\, G^{\mathrm{loc},>}_{jq}(t)\, w_{pqrs}(t)\, G^{\mathrm{loc},<}_{rk}(t)\,G^{\mathrm{loc},<}_{sl}(t) -(>\leftrightarrow <)\big\},
    \label{eq:def_Psi_loc}\\
    \Pi^\gtrless_{ijkl}(t,t') = -\mathrm{i}\hbar \sum_{pqr} \big[\mathcal{G}^\mathrm{loc}_{ipkq}(t)\,w_{qjpr}(t)\, G^\gtrless_{rl}(t,t')+ \big\{ G^{\mathrm{loc},<}_{ip}(t)\, w_{p q k r}(t) - w_{iq pr}(t)\,G^{\mathrm{loc},<}_{pk}(t) \big\}\, \mathcal{G}^\gtrless_{rjql}(t,t')\big],
    \label{eq:def_Pi}\\
    \Pi^\mathrm{loc}_{ijkl}(t) = -\mathrm{i}\hbar \sum_{pqr} \big[\mathcal{G}^\mathrm{loc}_{ipkq}(t) \big\{w_{qjpr}(t)\, G^{\mathrm{loc},<}_{rl}(t) - G^{\mathrm{loc},<}_{jr}(t)\,w_{qrpl}(t) \big\}+ \big\{ G^{\mathrm{loc},<}_{ip}(t)\,w_{pqkr}(t) - w_{iqpr}(t)\, G^{\mathrm{loc},<}_{pk}(t)\big\}\, \mathcal{G}^\mathrm{loc}_{rjql}(t) \big].
    \label{eq:def_Pi_loc}
\end{gather}
\end{widetext} 
For the special case $h^\mathrm{loc}=h^\mathrm{HF}$, i.e., mean-field reference propagators, Eqs.~\eqref{eq:EOM_G1_RTDE-MF}--\eqref{eq:def_Pi_loc} reduce to the $GW$ equations of motion derived in Ref.~\cite{reeves2024}. The generalization to an arbitrary time-local reference propagator $G^\mathrm{loc}$ affects the contributions to the collision term, the effective two-particle Hamiltonian, the equations of motion of the reference system, and the boundary conditions discussed below. The single-particle propagation of the full Green's function in Eq.~\eqref{eq:EOM_G1_RTDE-MF}, in contrast, is generated by $h^\mathrm{HF}$ for every choice of reference, since $h^\mathrm{HF}$ is the mean-field part of the physical self-energy, cf.~Eq.~\eqref{eq:KBE_HF}, and is evaluated with the time-diagonal Green's function fixed by the boundary conditions.

\begin{figure}[t]
\centering
\includegraphics[width=\columnwidth]{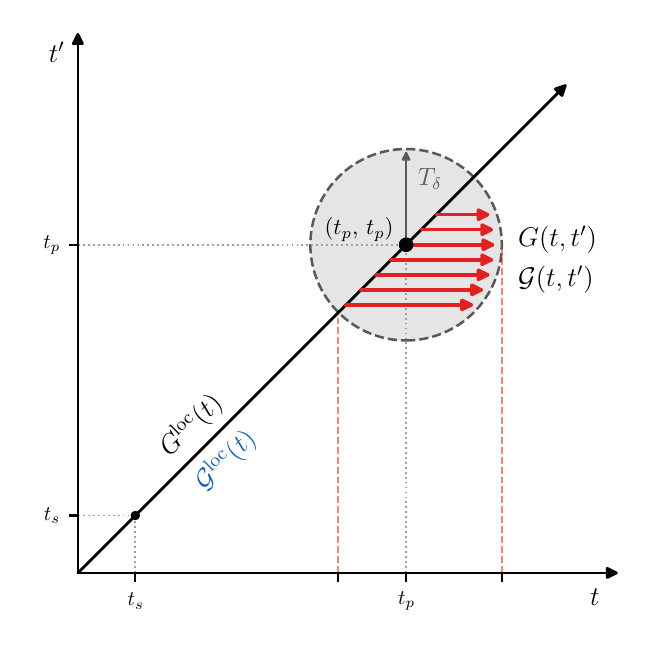}

\caption{Schematic of the off-diagonal reconstruction of the two-time
Green's functions in the $(t,t')$ plane. Propagation is first performed along
the time diagonal $t'=t$, yielding the local (time-diagonal) components
$G^{\mathrm{loc}}(t)$ and $\mathcal{G}^{\mathrm{loc}}(t)$. Within a probe window
(dashed circle) of radius $T_\delta$ centered on the probe time $(t_p,t_p)$, the
off-diagonal components $G(t,t')$ and $\mathcal{G}(t,t')$ are then reconstructed
by time stepping in $t$-direction, at fixed $t'$ (red arrows, $t'\le t\le T_{\max}$).
Because only the off-diagonal elements inside this window are required, the
overall cost is set by the time-diagonal propagation, while $T_\delta$ controls
the spectral resolution of the probe at $t_p$. $\mathcal{G}^\mathrm{loc}(t)$ only appears within the RT-DE scheme for self-energy approximations beyond SOA.}
\label{fig:rtde_schematic}
\end{figure}

The structure of Eqs.~\eqref{eq:G2_GW_MF_propagators_1} and \eqref{eq:G2_GW_MF_propagators_2} can be understood as follows  [for an illustration see Fig.~\ref{fig:rtde_schematic}]: The first terms involving $h^\mathrm{loc}$ describe the propagation determined by the time-local reference Hamiltonian, and the terms $\Psi^\gtrless$ as well as $\Psi^\mathrm{loc}$ constitute the source terms for the correlated two-particle Green's function and describe second-order scattering contributions. Further, $\Pi^\gtrless$ and $\Pi^\mathrm{loc}$ correspond to the iteration of the particle--hole bubble and thereby generate the resummation of the polarization diagrams to infinite order, leading to the buildup of dynamical screening. The SOA can be recovered by neglecting the polarization terms $\Pi$ and replacing the interaction tensor with the antisymmetric one, $w \rightarrow w^-$ in $\Psi$. As the time-local correlation function $\mathcal{G}^\mathrm{loc}(t)$ only enters Eq.~\eqref{eq:G2_GW_MF_propagators_1} via $\Pi$, this decouples the equation from the equation of motion for $\mathcal{G}^\mathrm{loc}(t)$, Eq.~\eqref{eq:G2_GW_MF_propagators_2}. The time-diagonal function $\mathcal{G}^\mathrm{loc}(t)$ follows from Eq.~\eqref{eq:G2_GW} by evaluating all Green's functions with $G^\mathrm{loc}$ and setting $t'=t$, i.e., $\mathcal{G}^\mathrm{loc}(t)= \mathcal{G}[G^\mathrm{loc},G^\mathrm{loc}](t,t)$. In general, this construction implies $\mathcal{G}^\mathrm{loc}(t)\neq \mathcal{G}^\gtrless(t,t)$, as $\mathcal{G}^\gtrless(t,t)$ still involves a self-consistent $G$, cf. Eq.~\eqref{eq:G2_GW}. 

Comparing Eqs.~\eqref{eq:G2_GW_MF_propagators_1} and \eqref{eq:G2_GW_MF_propagators_2}, the two equations differ in two aspects. First, the off-diagonal equation contains only three terms involving $h^\mathrm{loc}$, whereas the time-diagonal equation contains four. The reason is that $\partial_t$ does not act on the $t'$-dependence carried by $G_{sl}(\bar{t},t')$ inside Eq.~\eqref{eq:G2_GW}, so that the corresponding term is absent in Eq.~\eqref{eq:G2_GW_MF_propagators_1}. Second, the source and polarization terms differ in the propagator carrying the second time argument. In the off-diagonal case, $\Psi^\gtrless$ and $\Pi^\gtrless$ contain the full Green's function $G^\gtrless(t,t')$ and the correlated two-particle Green's function $\mathcal{G}^\gtrless(t,t')$, respectively, while in the time-diagonal case these are replaced by the corresponding reference quantities, $G^{\mathrm{loc},\gtrless}(t)$ and $\mathcal{G}^\mathrm{loc}(t)$. In addition, $\Pi^\mathrm{loc}$ contains one further contribution that has no counterpart in $\Pi^\gtrless$.

The equations of motion for the retarded and advanced components of the single-particle and two-particle Green's functions follow from the replacement $G^\gtrless \rightarrow G^\mathrm{R/A}$ in Eqs.~\eqref{eq:EOM_G1_RTDE-MF} and \eqref{eq:G2_GW_MF_propagators_1}.

\subsection{Boundary conditions} \label{ss:boundary_conditions}
An equivalent reformulation of the integral equation~\eqref{eq:G2_GW} into differential equations requires proper boundary conditions on the time diagonal. For the retarded and advanced components, these are given by
\begin{align}
    G^\mathrm{R}_{ij}(t,t) & =  \frac{1}{\mathrm{i}\hbar}\delta_{ij}
    = - G^\mathrm{A}_{ij}(t,t),\label{eq:GR_init_1}\\ 
    \mathcal{G}^\mathrm{R/A}_{ijkl}(t,t) &= 0.\label{eq:GR_init_2}
\end{align}
For the greater and lesser components, proper boundary conditions require considering Eq.~\eqref{eq:G2_GW} on the time diagonal. As the diagonal values, $G^\gtrless(t,t)$ and $\mathcal{G}^\gtrless(t,t)$, then have to be evaluated for every time $t$, this would lead to contributions coming from the derivative of the self-consistent Green's function, i.e., $\sim\int \chi(t,\bar t) w(\bar t)G^\mathrm{loc}(t,\bar t)\partial_t G(\bar t,t)\mathrm{d}\bar t$. This, in turn, would reintroduce the memory integrals in the form of $\sim \int \chi(t,\bar t) w(\bar t)G^\mathrm{loc}(t,\bar t) I(\bar t,t)\mathrm{d}\bar t$ and, thus, bring us back to the original scaling problem. In other words, the off-diagonal evolution away from the diagonal is generated by time-local Hamiltonians, but the equal-time value of $G^\gtrless(t,t)$, at each $t$, is itself a fully correlated observable, whose exact evolution requires the complete memory that cannot be eliminated in favor of coupled time-local equations for $G$ and $\mathcal{G}$. Hence, it is necessary to employ a further approximation for the boundary conditions. In line with the treatment of the reference propagator, all correlation contributions on the time diagonal are neglected, and we set
\begin{align}
    G^\gtrless(t,t) &= G^{\mathrm{loc},\gtrless}(t),
    \label{eq:BC_G}\\
        \mathcal{G}^\gtrless(t,t) &=0.
    \label{eq:BC_G2}
\end{align}
This choice is the time-domain counterpart of the one-shot approximation itself, as correlations are added in the off-diagonal direction on top of a reference that also supplies the diagonal data. As a consequence of this, time-diagonal observables, such as densities and occupations, are described at the level of the reference propagator, while correlations enter only in the off-diagonal direction, i.e., in the spectral information. An improved treatment of the time diagonal becomes possible if the reference propagator itself contains correlations, as is the case for the GKBA-based scheme discussed in Sec.~\ref{ss:HFGKBA}, where the boundary conditions~\eqref{eq:BC_G} and \eqref{eq:BC_G2} are replaced by correlated ones.

\subsection{Computational scaling}
Let us briefly consider the computational scaling. The equations of motion for $G^\gtrless(t,t')$ and $\mathcal{G}^\gtrless(t,t')$ are propagated with respect to $t$, for fixed $t'$, starting from the boundary conditions on the time diagonal. Each off-diagonal propagation is thus linear in the number of time steps. Moreover, for the computation of the time-resolved spectral function, cf.~Sec.~\ref{ss:spectral_function}, only the $N_{t,\delta}$ time steps within the probe window are required [cf. Fig.~\ref{fig:rtde_schematic}], so that at most $N_{t,\delta}$ off-diagonal propagations of length $N_{t,\delta}$ have to be performed, in addition to the time-diagonal propagation of the reference system over $N_t$ steps. The total runtime, therefore, scales as $\mathcal{O}(N_t + N_{t,\delta}^2)$, i.e., linearly in the total propagation time, with usually $N_{t,\delta}\ll N_t$ and $N_{t,\delta}$ being fixed by the probe duration $\delta$ and effectively independent of $N_t$. The number of required off-diagonal propagations can be reduced further by interpolation techniques \cite{reeves2025}. Since the individual off-diagonal propagations are independent of each other, they can be performed in parallel, which reduces the runtime at the expense of holding several time slices in memory simultaneously, i.e., by increasing the memory. If they are instead performed sequentially, only a single time slice has to be stored at any time, and the memory consumption, which is dominated by the four-index functions $\mathcal{G}^\gtrless_{ijkl}(t,t')$ of size $\mathcal{O}(N_b^4)$, is independent of the number of time steps. The scaling with the basis size is determined by the contractions in Eq.~\eqref{eq:G2_GW_MF_propagators_1} and heavily depends on the structure of the interaction. For a general tensor $w_{ijkl}$, the polarization terms $\Pi$ contain the contraction of two four-index quantities over two indices, giving $\mathcal{O}(N_b^6)$ per time step. For density--density interactions, $w_{ijkl}=v_{ij}\delta_{ik}\delta_{jl}$, as considered in this work, the scaling reduces to $\mathcal{O}(N_b^5)$. Moreover, for the Hubbard model with purely on-site interaction and a sparse single-particle Hamiltonian, the cost reduces to $\mathcal{O}(N_b^4)$. For spatially homogeneous systems in the momentum basis, symmetry reduces the memory to $\mathcal{O}(N_b^3)$ and the runtime to at most $\mathcal{O}(N_b^4)$ \cite{joost_prb_20}.

\section{Time-local reference propagators}\label{s:reference_propagators}
As in the equilibrium one-shot approaches, there is considerable freedom in the choice of the reference propagator. The elimination of the memory integrals constrains only the time-off-diagonal propagation, which has to be generated by a time-local Hamiltonian $h^\mathrm{loc}$ without a collision term, cf.~Eq.~\eqref{eq:EOM_Gloc}. The propagation along the time diagonal is not restricted in this way and may include correlation contributions. Within this constraint we consider three references, ordered by the degree of correlation they contain: mean-field (Hartree--Fock) propagators (Sec.~\ref{ss:MF_propagators}), statically screened propagators (Sec.~\ref{ss:HSEX_propagators}), and propagators based on the HF-GKBA (Sec.~\ref{ss:HFGKBA}). Throughout, $GW[\mathrm{X}]$ denotes the RT-DE with the $GW$ self-energy evaluated with the reference propagator ``$\mathrm{X}$''. $GW$[HF] corresponds to the mean-field-referenced scheme derived in Ref.~\cite{reeves2024}, whereas $GW$[HSEX] and $GW$[GKBA] are introduced in this work.
\subsection{Mean-field propagators: HF} \label{ss:MF_propagators}
The standard choice for one-shot corrections is to consider either the ideal ($G^0$) or the mean-field ($G^\mathrm{MF}$) Green's function as reference. This is also the choice made in the original formulation of the RT-DE~\cite{reeves2024} and in all subsequent applications~\cite{reeves2025, blommel2026}, and it serves as the reference case against which the propagators introduced in the following subsections are compared. Here, we consider the mean-field Green's function, i.e., we use the effective time-local single-particle Hartree--Fock Hamiltonian,
\begin{align}
    h^\mathrm{loc}=h^\mathrm{HF}.
\end{align}
This choice corresponds to a time-dependent Hartree--Fock treatment of the reference system and constitutes the nonequilibrium analog of a $G_0W_0$ calculation on top of a Hartree--Fock starting point. 
\subsection{Statically screened mean-field propagators: HSEX}
\label{ss:HSEX_propagators}
The mean-field reference of Sec.~\ref{ss:MF_propagators} treats the reference system with the bare interaction. In systems with appreciable screening, however, the effective interaction entering the exchange channel is significantly reduced, thus affecting the dynamics of the system. A natural refinement of the reference that retains its time-local (memory-free) structure is, therefore, to replace the bare interaction in the exchange self-energy by a screened one, giving a Hartree plus screened exchange (HSEX) reference. Such screened-exchange references are well established as starting points in equilibrium many-body perturbation theory, but have not been used as reference propagators for the RT-DE.

The screened interaction is generated from a Hartree--Fock trajectory that is propagated alongside the HSEX reference. At each time $t$, its instantaneous Hamiltonian is diagonalized, $h^\mathrm{HF}(t) =U(t)\,\mathrm{diag}\big(\varepsilon_\alpha(t)\big)U^\dagger(t)$, and the occupations $f_\alpha(t)$ are obtained by projecting the time-diagonal Hartree--Fock Green's function $G^\mathrm{HF}(t)$ onto the instantaneous eigenbasis. The static limit of the retarded polarizability then reads
\begin{align}
    \chi^{0,\mathrm{R}}_{ijkl}(\omega\to 0,t) =\sum_{\alpha\beta}&\,\bigg(
    U_{i\alpha}(t)U_{j\beta}(t)U^*_{k\beta}(t)U^*_{l\alpha}(t)\,\\&\times
    \frac{f_\alpha(t) - f_\beta(t)}
         {\varepsilon_\alpha(t) - \varepsilon_\beta(t) + \mathrm{i}\eta}\bigg),
    \label{eq:chi0_static}
\end{align}
where $\eta$ is a small regularization parameter (in this work it is of the order $10^{-6}$). The screened interaction is then obtained as
\begin{align}
    W(t) =\big[1 - w(t)\chi^{0,\mathrm{R}}(\omega\rightarrow 0,t)\,\big]^{-1} w(t). \label{eq:W_hsex}
\end{align}
The time-local static screening refers here to the frequency dependence of $W$, which is evaluated in the limit $\omega\to 0$, and not to its time dependence as $W(t)$ is rebuilt at every time step. During the excitation the pump-induced carriers modify the screening, so that the HSEX reference incorporates a nonequilibrium, time-dependent screening at the mean-field level while remaining memory-free. Constructing $W(t)$ from the HSEX trajectory itself instead of the parallel Hartree--Fock one changes the results only marginally for the systems considered here (not shown).

The HSEX reference Hamiltonian is given by
\begin{align}
    h^\mathrm{HSEX}(t) = h(t) + \Sigma^\mathrm{H}(t)  + \Sigma^\mathrm{SEX}(t),
    \label{eq:h_HSEX}
\end{align}
where the Hartree term, $\Sigma^\mathrm{H}$, is evaluated with the bare interaction, and the screened-exchange term with the instantaneous $W(t)$,
\begin{align}
    \Sigma^\mathrm{SEX}_{ij}(t) = \mathrm{i}\hbar\sum_{pq}
    W_{ipqj}(t)\, G^<_{qp}(t).
    \label{eq:sigma_sex}
\end{align}
For the density--density interaction considered in the following, the same-spin interaction of two electrons in the same orbital vanishes identically due to the Pauli principle, so that the bare exchange self-energy has a vanishing diagonal. The exact static exchange self-energy has no on-site contribution in the basis in which the interaction is local. Replacing the bare interaction by $W$ does not preserve this property, since the screened interaction acquires a nonvanishing on-site element even where the bare one vanishes. The diagonal of $\Sigma^\mathrm{SEX}$ then describes the interaction of a particle with its own density, and removing it restores the zero required by antisymmetry. We will therefore consider the following \textit{nonstandard} modification of the HSEX self-energy for density--density interactions in the orbital basis given by
\begin{align}
    \tilde{\Sigma}^\mathrm{SEX}_{ij}(t)= (1-\delta_{ij})\Sigma^\mathrm{SEX}_{ij}(t).  \label{eq:sigma_sex_modification}
\end{align}
In the following, we will only use this modified version and simply refer to it as HSEX. 

An analogous deficiency is known for the $GW$ self-energy itself, where the particle contributes to the polarizability determining $W$ and, therefore, interacts with the polarization it induces. Removing this self-polarization error requires exchange-type vertex corrections, as in second-order screened exchange (SOSEX)~\cite{Gruneis2009, Romaniello2009,Nelson2007}. Such corrections are dynamical and would reintroduce the memory dependence that the time-local construction is designed to avoid, which is why the diagonal removal in Eq.~\eqref{eq:sigma_sex_modification} serves as a static substitute here. The procedure is tied to the orbital basis in which the interaction is local and is not intended as a general construction.

It is worth noting how the approach presented here differs from the way screened-exchange references are commonly employed in \textit{ab initio} real-time calculations, in two respects. First, the screened interaction there is evaluated once for the equilibrium ground state and kept fixed during the propagation. The time-diagonal Green's function entering the self-energy is propagated, but $W$ is not~\cite{attaccalite2011,attaccalite2013}. The carrier redistribution induced by the pump changes the polarizability and hence the screening, and this feedback has been shown to affect the nonequilibrium dynamics~\cite{perfetto2020}. Consequently, this approximation is only valid for very weak excitations. Second, the HSEX self-energy usually enters only through its change relative to equilibrium, $\Delta\Sigma^\mathrm{HSEX}(t) = \Sigma^\mathrm{HSEX}(t) - \Sigma^\mathrm{HSEX}_\mathrm{eq}$, added on top of a Hamiltonian that already carries the equilibrium quasiparticle structure from a separate calculation, for instance at the DFT or $GW$ level~\cite{attaccalite2011,perfetto2020}. The two roles are thereby assigned to different methods: the equilibrium structure is calculated elsewhere, and the screened-exchange term accounts only for the response to the external field. It is then never required to reproduce absolute quasiparticle energies. Any deficiency already present in equilibrium is subtracted together with $\Sigma^\mathrm{HSEX}_\mathrm{eq}$, and only its change under excitation survives. In the present work no such division exists as $h^\mathrm{HSEX}$ is self-contained and alone determines the ground state, the quasiparticle structure and the dynamics, so that a spurious on-site contribution enters the spectrum directly and must be removed explicitly. Finally, in the limiting case where the screened interaction is replaced by a fixed, globally screened one, the screened-exchange term takes the same form as the nonlocal exchange of a range-separated hybrid functional, with the screening parameter playing the role of the range separation \cite{heyd_hse_2003,bylander_kleinman_1990}.

Common to both is that the spectral information carried by $\Sigma^\mathrm{HSEX}(t)$ alone is that of a time-local self-energy. Its instantaneous eigenvalues give one pole per orbital, without lifetime broadening or satellite structures, so that dynamical correlation effects are absent from the spectrum.
\subsection{GKBA propagators}
\label{ss:HFGKBA}
The generalized Kadanoff--Baym ansatz (GKBA) \cite{lipavsky_generalized_1986} reconstructs the time-off-diagonal lesser and greater components of the Green's function from the time-diagonal ones, using the retarded and advanced components,
\begin{align}
    G^{\mathrm{GKBA},\gtrless}(t,t')  = & \mathrm{i}\hbar \big\{
    G^{\mathrm{GKBA,R}}(t,t')\, G^{\mathrm{GKBA},\gtrless}(t')
    \\ & - G^{\mathrm{GKBA},\gtrless}(t)\, G^{\mathrm{GKBA,A}}(t,t') \big\}. \quad
    \label{eq:GKBA}
\end{align}
The retarded and advanced Green's functions are not available without further approximation and are commonly approximated at the mean-field level, which gives rise to the so-called Hartree--Fock GKBA (HF-GKBA),
\begin{align}
    \mathrm{i}\hbar \partial_t G^{\mathrm{GKBA,R}}(t,t') =& \delta(t-t')  +   \label{eq:EOM_GR_GKBA}
\\& h^{\mathrm{HF}}\big[G^\mathrm{GKBA}\big](t)\, G^{\mathrm{GKBA,R}}(t,t').
\end{align}
The Green's function $G^\mathrm{GKBA}$ entering the effective single-particle Hamiltonian is the correlated one on the time diagonal, which is propagated including the collision term. The HF-GKBA therefore combines a correlated time diagonal with an off-diagonal reconstruction at mean-field level, into which correlations enter only indirectly, through the time-diagonal Green's function in $h^{\mathrm{HF}}$.

Replacing the two-time correlated KBE by Eq.~\eqref{eq:GKBA} reduces the cubic scaling with the number of time steps to $\mathcal{O}(N_t^2)$, for the second-order Born approximation, where the collision integral involves a single time integration. For self-energies that resum entire diagram classes, such as $GW$, this reconstruction, however, does not reduce the numerical scaling. Significant progress is achieved with the G1--G2 scheme \cite{schluenzen_prl_20,joost_prb_20} that is a time-local exact reformulation of the HF-GKBA and has time-linear scaling for all relevant self-energy approximations, including second-order Born, $GW$, $T$-matrix, and the dynamically screened ladder approximation \cite{joost_prb_22}. The G1--G2 scheme eliminates the time memory integral in favor of two coupled time-local equations for the  time-diagonal functions $G^{\mathrm{GKBA},\gtrless}(t)$ and the correlated part $\mathcal{G}^{\mathrm{GKBA}}(t)$ of the two-particle Green's function.

Another issue concerns the quality of the resulting approximation. For small Hubbard clusters it has been found that full two-time KBE solutions with standard self-energy approximations exhibit an artificial damping of the dynamics towards unphysical steady states~\cite{von_friesen_successes_2009,PuigvonFriesen2010}, an artifact that the HF-GKBA avoids~\cite{hermanns_hubbard_2014, schluenzen_cpp16}. Moreover, the HF-GKBA describes scattering-induced population transfer, which mean-field approximations cannot capture by construction. Its spectral information, in contrast, remains at the mean-field level. By Eq.~\eqref{eq:EOM_GR_GKBA} the two-time dynamics are generated by $h^{\mathrm{HF}}$ alone, so that the spectrum consists of peaks at the mean-field eigenvalues, without quasiparticle renormalization or satellite structures. The weights of these peaks, however, are the occupations of the correlated time-diagonal Green's function. The same single-particle energies therefore contribute to the occupied and to the unoccupied spectrum, and GKBA spectra may contain artificial excitations within an otherwise mean-field-like level structure.

Although the HF-GKBA involves collision contributions on the time diagonal, it is nevertheless an admissible reference for the RT-DE. This is possible because the derivation of the RT-DE constrains only the time-off-diagonal reference propagator. The propagation along the time diagonal is not subject to this restriction and may contain correlation contributions and memory. This is a slight generalization of the time-local reference introduced in Sec.~\ref{ss:RTDE_definition}.

Combining the G1--G2 scheme and RT-DE is a natural step as the two approaches are based on the same structural reformulation of the KBE and work with an analogous set of quantities---the single-particle Green's function $G$ and the correlated part of the two-particle Green's function $\mathcal{G}$.  Moreover, within the $GW$ approximation, the RT-DE already requires the time-diagonal correlated part $\mathcal{G}^{\mathrm{loc}}(t)$, cf.~Eq.~\eqref{eq:def_Pi}, because it enters the polarization term $\Pi^\gtrless$. Consequently, no additional quantity has to be propagated or stored, and the additional cost amounts to one contraction per time step.

Accordingly, the equations of motion are those of the mean-field propagators, Eqs.~\eqref{eq:G2_GW_MF_propagators_1} and \eqref{eq:G2_GW_MF_propagators_2}, with the replacements $G^{\mathrm{loc}}(t)\rightarrow G^{\mathrm{GKBA}}(t)$ and $h^{\mathrm{loc}}\rightarrow h^{\mathrm{HF}}[G^{\mathrm{GKBA}}]$, where, in addition, the equation of motion for $G^{\mathrm{GKBA},\gtrless}(t)$ contains the collision term,
\begin{align}
    \mathrm{i}\hbar \partial_t G^{\mathrm{GKBA},<}(t) = h^{\mathrm{HF}}(t)\, G^{\mathrm{GKBA},<}(t) + \mathcal{I}(t) + \mathrm{h.c.},
\end{align}
with the time-diagonal collision term
\begin{align}
    \mathcal{I}_{ij}(t) = -\mathrm{i}\hbar \sum_{pqr} w_{ipqr}(t)\,
    \mathcal{G}^{\mathrm{GKBA}}_{rqpj}(t).
\end{align}
For the off-diagonal propagation we use the correlated boundary conditions
\begin{align}
    G^\gtrless(t,t) &= G^{\mathrm{GKBA},\gtrless}(t),\label{eq:BC_G_GKBA}\\
    \mathcal{G}^\gtrless(t,t) &= \mathcal{G}^{\mathrm{GKBA}}(t), \label{eq:BC_G2_GKBA}
\end{align}
so that correlations enter both the time-diagonal observables and the off-diagonal spectral information. This does not resolve the boundary-value problem of the RT-DE. Equations~\eqref{eq:BC_G_GKBA} and \eqref{eq:BC_G2_GKBA} remain approximate, since the exact diagonal values of $G^\gtrless$ and $\mathcal{G}^\gtrless$ would again require the memory integrals discussed in Sec.~\ref{ss:RTDE_definition}. The resulting scheme, denoted $GW$[GKBA] in the following, is thus of \textit{semi-self-consistent} character.

\section{Numerical Results} \label{s:numerics}

We now assess the RT-DE within the $GW$ approximation for the reference propagators  introduced in Sec.~\ref{s:reference_propagators}, for small systems against exact results  that are obtained by exact diagonalization (ED) simulations. Based on the benchmarks for small systems, we then move on to larger systems, where no exact data are available. Only for larger systems, and thus finer momentum resolution, can the excitonic replicas and satellites of the photoexcited state be resolved as dispersing branches (Sec.~\ref{ss:ns40}). We compare to mean-field approaches, HF and HSEX, and the RT-DE within the second-order Born approximation (SOA) using HF propagators. For simplicity we refer to this approximation as SOA instead of SOA[HF] in the following. SOA is included in the small-system comparisons of Secs.~\ref{ss:eq_benchmark}--\ref{ss:undoped}, whereas the large-system section is restricted to the $GW$ schemes and their references.

\subsection{Model and setup} \label{ss:model}
Two-band models with an interband Coulomb interaction are the standard minimal description of gapped systems in which electron--hole correlations dominate the low-energy physics. In particular, they underlie the semiconductor Bloch equations~\cite{haug-koch-book} and their nonequilibrium Green's functions extensions, with which the ultrafast dynamics of optically excited electron--hole plasmas have been studied extensively, see, e.g., Ref.~\cite{kwong-etal.98pss} and references therein.

Here we consider a two-band extended Hubbard chain with $N_s$ sites and periodic boundary conditions, with a valence ($v$) and a conduction ($c$) orbital per site,
\begin{align}
    \hat{H}(t) =\,& \sum_{\alpha\beta}\sum_{ij,\sigma} h^{\alpha\beta}_{ij}(t)\hat{c}^\dagger_{i\alpha\sigma}\hat{c}_{j\beta\sigma} + U\sum_{i,\alpha}\hat{n}_{i\alpha\uparrow}\hat{n}_{i\alpha\downarrow}
    \notag\\
    &+ U\sum_{i}\hat{n}_{ic}\hat{n}_{iv} + \gamma U\sum_{\alpha\beta}\sum_{i<j} \frac{\hat{n}_{i\alpha}\,\hat{n}_{j\beta}}{d(i,j)}, \label{eq:model_H}
\end{align}
where $i,j$ denote the site, $\alpha,\beta$ the band and $\sigma$ the spin index, and $\hat{n}_{i\alpha}=\sum_\sigma \hat{n}_{i\alpha\sigma}$ is the band-resolved local density. The single particle part is given by
\begin{align}
    h^{\alpha\beta}_{ij}(t) =\,& (-1)^{\delta_{\alpha c}}J \delta_{\alpha\beta}\delta_{\langle i,j\rangle} + \varepsilon_\alpha\delta_{\alpha\beta}\delta_{ij} \notag\\
    &+ E(t)\,(1-\delta_{\alpha\beta})\delta_{ij},
    \label{eq:model_h1}
\end{align}
where $J$ denotes the hopping constant between neighboring sites on the same band and $E(t)$ is the external field coupling the different bands. Further, $\delta_{\langle i,j\rangle}$ is one if $i$ and $j$ are nearest neighbors and zero otherwise.

In the following we use units with $\hbar=J=a=1$, where $a$ is the lattice constant, so that energies are measured in units of $J$, frequencies in units of $J/\hbar$ and times in units of $\hbar/J$. Moreover, distances are measured in units of $a$ and momenta in units of $\hbar/a$.

The two bands have equal hopping magnitude and opposite curvature, each with a bandwidth of $4$. In the noninteracting system they are separated by $\varepsilon_c-\varepsilon_v-4$, at $k=\Gamma$, where the valence band maximum and the conduction band minimum coincide, and by $\varepsilon_c-\varepsilon_v$, at $k=\pm\pi/2$. The interaction opens the gap further so that at the parameters used below, the Hartree--Fock quasiparticle gap of the undoped system is $E_\mathrm{G}=6.5$, with the band edges at $\mp3.25$ relative to the chemical potential.  For the undoped system this gap is exact: since the interaction conserves the band occupations separately, the sectors with a single valence hole or a single conduction electron are spanned by single-particle configurations only, so that the exact addition and removal energies coincide with the Hartree--Fock eigenvalues. It is also the equilibrium gap of all three references, as neither the static screening of the HSEX reference nor the collision term of the GKBA reference is active in the filled two-band ground state, cf.~Secs.~\ref{ss:HSEX_propagators} and \ref{ss:HFGKBA}.

The interaction consists of an on-site Hubbard term $U$, taken equal for the intra- and interband contributions, and a long-range Coulomb-type interaction of relative strength $\gamma$. Given periodic boundary conditions, the distance entering the latter is calculated as $d(i,j)=\min\{|i-j|,\,N_s-|i-j|\}$. A physically relevant generalization of this interaction in the two-band model is to consider different inter- and intraband interactions $U_\mathrm{inter}$ and $U_\mathrm{intra}$. We will, however, refrain from considering such an extension in this work.

The external field $E(t)$ couples the two bands and drives interband transitions. We consider a Gaussian-enveloped dipole pulse,
\begin{align}
    E(t) = E_0\,\cos\big[\omega_0(t-t_0)\big]\,e^{-\frac{(t-t_0)^2}{2T_0^2}}, \label{eq:pulse}
\end{align}
centered at $t_0=10$ and with width $T_0=2$. Since the electric field term in Eq.~\eqref{eq:model_h1} is diagonal in the site index, and, therefore, also in $k$, it induces only vertical interband transitions. The carrier frequency is chosen according to the parameters of the system to address the lowest vertical transition of the respective system (specified below). Pump amplitudes up to $E_0=0.5$ are considered, corresponding to excited-carrier fractions of up to about $45\%$. The largest amplitudes exceed experimentally realistic excitation densities and serve to probe the methods over a broad range of conditions.

Unless stated otherwise we fix $\varepsilon_c=-\varepsilon_v=4$, $U=2.5$ and $\gamma=0.5$. All energies are considered relative to the chemical potential $\mu$ of the exact ground state. 

Two fillings are considered in the following. The \emph{undoped} system has a completely filled valence band ($N_\uparrow=N_\downarrow=N_s$), corresponding to a gapped semiconductor with $\mu$ in the middle of the gap. The \emph{doped} system has a valence band that is $75\%$ filled and an empty conduction band. For the latter the interband quasiparticle gap remains open, but $\mu$ lies inside the valence band. For the calculations that are tested against exact benchmarks, we use $N_s=4$, while the large-system calculations of Sec.~\ref{ss:ns40} use $N_s=40$.

For the undoped system, as the valence band is fully occupied and the conduction band is empty, the ground state is exactly described by the Hartree--Fock solution, and any feature appearing at finite $E_0$ is of purely nonequilibrium origin. The doped system, in contrast, has a partially filled valence band and a correlated ground state already in equilibrium, which makes it the more demanding case for the equilibrium benchmark of Sec.~\ref{ss:eq_benchmark}.

\begin{figure}[tb]
  \includegraphics[width=\columnwidth]{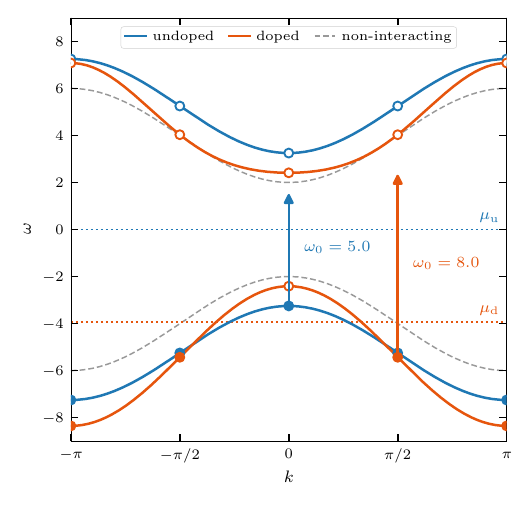}
  \caption{%
  Exact quasiparticle band structure of the two-band chain with four sites at $U=2.5$ and $\gamma=0.5$, given by the position of the dominant pole of the exact removal spectrum (of the addition spectrum for empty states), for the \textit{undoped} system (blue) with fully filled valence band and the \textit{hole-doped} system (orange) with partially filled valence band (no particles at $k=\Gamma$). For the undoped system these energies coincide with the Hartree--Fock eigenvalues, whereas for the doped system the approximate schemes yield band structures that differ from the exact one, cf.~Fig.~\ref{fig:gs_spectra}. Gray dashed lines are the shared noninteracting bands. Filled/open circles mark occupied/empty single-particle states at $k=-\pi,-\pi/2,0,\pi/2,\pi$. The open orange circle at $\Gamma$ is the doping hole. Dotted lines give the chemical potentials: $\mu_\mathrm{u}$ lies in the mid-gap (undoped), while $\mu_\mathrm{d}$ sits inside the valence band (doped). Vertical arrows are the pump photon energies $\omega_0=5.0$ (undoped, at $\Gamma$) and $8.0$ (doped, at $\pi/2$), each tuned near the respective optical gap. For the undoped system at $N_s=40$ (Sec.~\ref{ss:ns40}) the same bands are sampled on a correspondingly denser grid (not marked).}
  \label{fig:band_structure}
\end{figure}

Figure~\ref{fig:band_structure} shows the exact interacting quasiparticle band structures of both systems, together with the pump photon energies used below. For the undoped system all valence states are occupied and the lowest vertical transition is the one at $k=\Gamma$. In the doped system the valence state at $\Gamma$ carries the doping hole and is not available, so that the lowest allowed transition is shifted to $k=\pm\pi/2$. The frequency of the excitation, $\omega_0$, is set slightly below the quasiparticle gap, at the optical gap of the system, i.e., close to the energy of the lowest bound electron--hole excitation rather than to the energy required to create an unbound pair. For the undoped system this gives $\omega_0=5.0$ against a Hartree--Fock quasiparticle gap of $E_\mathrm{G}=6.5$, which is independent of the system size. The difference between the two energies is of the order of the exciton binding energy. Driving below the quasiparticle gap ensures that the excitation creates bound electron--hole pairs, which is the prerequisite for the excitonic features analyzed in Sec.~\ref{ss:ns40}. For the doped system, a frequency of $\omega_0 =8.0$ is chosen analogously.

All calculations are initialized in the noninteracting (zero-temperature) ground state with sharp occupations, and the interacting initial state is prepared by adiabatic switching of the interaction (see Ref.~\cite{hermanns_hubbard_2014, schluenzen_cpp16} for more details). The reference propagator and the time-diagonal correlated two-particle function $\mathcal{G}^\mathrm{loc}(t)$ are switched on simultaneously, starting from $\mathcal{G}^\mathrm{loc}(t_s)=0$ with $t_s=-50$ for the GKBA-based schemes, whose correlated time diagonal requires the slower switching, and $t_s=-25$ for all others, cf. Fig.~\ref{fig:rtde_schematic}. For the GKBA reference the two equations are, in addition, coupled through the collision term.

\subsection{Observables} \label{ss:numerics_observables}
The central observable is the time-resolved spectral function of Sec.~\ref{ss:spectral_function}, $A(\omega,t_p)\equiv A_\delta(\omega,t_p)$ (we drop the subscript $\delta$ below), evaluated at probe time $t_p$ with probe duration $\delta$. Parameter scans are performed at $\delta=3$, while individual spectra are displayed at $\delta=5$. The lower resolution is used for the scans and the error calculations, as the spectra are dominated by their overall structure, so that the comparison of the methods is governed by the gross distribution of spectral weight rather than by small displacements of individual peaks. The
sharper probe is used where individual peaks are to be identified.

With the probe of Eqs.~\eqref{eq:A_lesser} and \eqref{eq:A_greater}, the
frequency-integrated spectra obey the sum rules $\int A^<\,\mathrm{d}\omega = (\sqrt{\pi}/\delta)\,N_\mathrm{occ}$ and $\int A^>\,\mathrm{d}\omega = (\sqrt{\pi}/\delta)\,N_\mathrm{unocc}$, where $N_\mathrm{occ/unocc}$ denotes the number of occupied/unoccupied states. Upon frequency integration, Eqs.~\eqref{eq:A_lesser} and \eqref{eq:A_greater} reduce to the equal-time Green's functions, so that the sum rules depend only on the time-diagonal values. In the RT-DE these are fixed by the boundary conditions to those of the reference, cf.~Secs.~\ref{ss:boundary_conditions} and \ref{ss:HFGKBA}, which conserves the particle number in all cases considered. The sum rules are accordingly fulfilled by the numerical spectra to better than $10^{-5}$. Spectra are displayed in units of $\sqrt{\pi}/\delta$, so that $\int A^<\,\mathrm{d}\omega = N_\mathrm{occ}$ and $\int A^>\,\mathrm{d}\omega = N_\mathrm{unocc}$.

The deviation of an approximate spectrum from the exact one is quantified by the relative error
\begin{align}
    \epsilon_\mathrm{rel} = \frac{\big\lVert A^\gtrless_\mathrm{exact}- A^\gtrless_\mathrm{approx}\big\rVert_1}
    {\big\lVert A^\gtrless_\mathrm{exact}\big\rVert_1},
    \qquad
    \lVert A\rVert_1 = \int_{-\infty}^{\infty} |A(\omega)|\,\mathrm{d}\omega.
    \label{eq:eps_rel}
\end{align}
As a global norm, $\epsilon_\mathrm{rel}$ is dominated by the largest features of the spectrum, which are the coherent quasiparticle peaks. Where the weight transferred to the excited region is of interest, we resolve the error by spectral region, as described at the end of this section.
For the driven system, we additionally consider the population of excited states. From the time diagonal, we obtain the excited-carrier fraction
\begin{align}
    n_\mathrm{exc}=\frac{\Delta N_c}{N_\mathrm{el}} ,
\end{align}
where $\Delta N_c$ is the change of the conduction-band population, evaluated at the probe time $t_p$ from the $k$-resolved densities, and $N_\mathrm{el}$ the total electron number. Since the Hamiltonian (without the laser field) conserves the band occupations separately, $N_c$ is constant after the pulse.
From the spectral function, we define the cumulative excited spectral weight
\begin{align}
    f_\mathrm{exc}(\omega) = \frac{1}{N_\mathrm{occ}} \int_\mu^{\omega} A^<(\omega')\,\mathrm{d}\omega',
    \quad f_\mathrm{exc} = f_\mathrm{exc}(\omega\rightarrow\infty),  \label{eq:fexc}
\end{align}
i.e., $f_\mathrm{exc}$ is the fraction of occupied spectral weight above the chemical potential. In contrast to $n_\mathrm{exc}$, it differentiates the RT-DE schemes from their references.
Finally, the placement of the excited weight is quantified by $\Delta_\mathrm{exc}$, the error of the renormalized cumulative $\tilde{f}_\mathrm{exc}(\omega) = f_\mathrm{exc}(\omega)/f_\mathrm{exc}$ with respect to the exact result,
\begin{align}
    \Delta_\mathrm{exc}= \big\lVert \tilde{f}^\mathrm{exact}_\mathrm{exc}- \tilde{f}^\mathrm{approx}_\mathrm{exc}\big\rVert_1 .
\end{align}
The renormalization by $f_\mathrm{exc}$ makes $\Delta_\mathrm{exc}$ sensitive to the position of the excited weight only, complementary to its integrated amount $f_\mathrm{exc}$.
\subsection{Equilibrium doped system} \label{ss:eq_benchmark}
\begin{figure}[t]
  \includegraphics[width=\columnwidth]{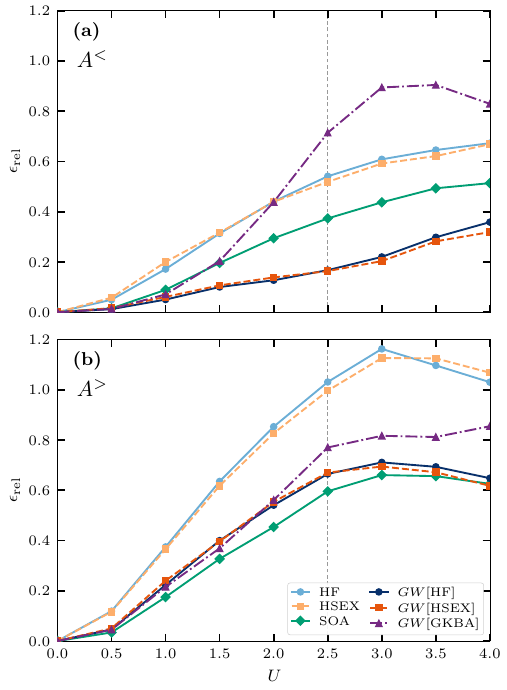}
  \caption{%
  Relative error $\epsilon_{\mathrm{rel}}$, Eq.~\eqref{eq:eps_rel}, of the
  ground-state spectral function for the doped system with four sites for varying on-site interaction $U$ compared to the exact result, for (a) the occupied part $A^<$ and (b) the unoccupied part $A^>$, as $U$ is scanned from $0$ to $4$ (probe time $t_p=15$, $\delta=3$). The dashed vertical line marks the $U=2.5$ working point used in the nonequilibrium applications. The long-range interaction parameter is fixed $\gamma=0.5$.}
  \label{fig:gs_uscan}
\end{figure}

We first benchmark the correlated ground state. Only the doped system is considered here, as the undoped ground state can trivially be described by Hartree--Fock. Figure~\ref{fig:gs_uscan} shows the relative error $\epsilon_\mathrm{rel}$ of the occupied [$A^<$, panel~(a)] and unoccupied [$A^>$, panel~(b)] spectral functions as the on-site interaction is scanned from $U=0$ to $4$.

\begin{figure}[tb]
  \includegraphics[width=\columnwidth]{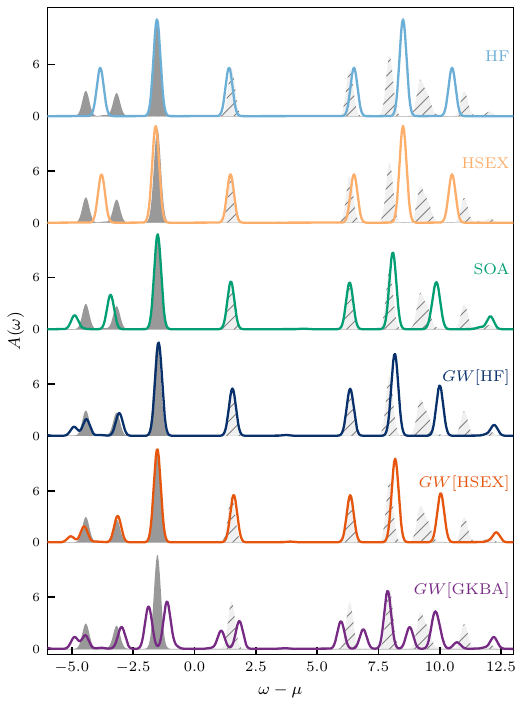}
 \caption{%
    Doped ground-state spectral function $A$ of the four-site chain at $U=2.5$. The exact result is drawn behind every trace: gray fill for the occupied $A^<$ and hatched fill for the unoccupied $A^>$. All spectra satisfy $\int A^<=N_{\mathrm{occ}}=6$, $\int A^>=N_\mathrm{unocc}=10$, $\int A=16$ (in units of $\sqrt{\pi}/\delta$). Spectra are displayed at $\delta=5$.}
    \label{fig:gs_spectra}
\end{figure}

Three groups of methods emerge for the occupied spectrum. The $GW$ schemes with uncorrelated references, $GW$[HF] and $GW$[HSEX], are the most accurate over the entire $U$ range and remain close to one another throughout. They are followed by SOA, while the mean-field approximations HF and HSEX are the least accurate. The explicit spectra at the working point $U=2.5$, Fig.~\ref{fig:gs_spectra}, reveal the origin of this ordering. $GW$[HF] and $GW$[HSEX] reproduce all peaks of the exact spectrum, including the incoherent correlation satellites, whereas the mean-field approximations capture only the coherent quasiparticle peaks.
\begin{figure*}[tb!]
  \includegraphics[width=\textwidth]{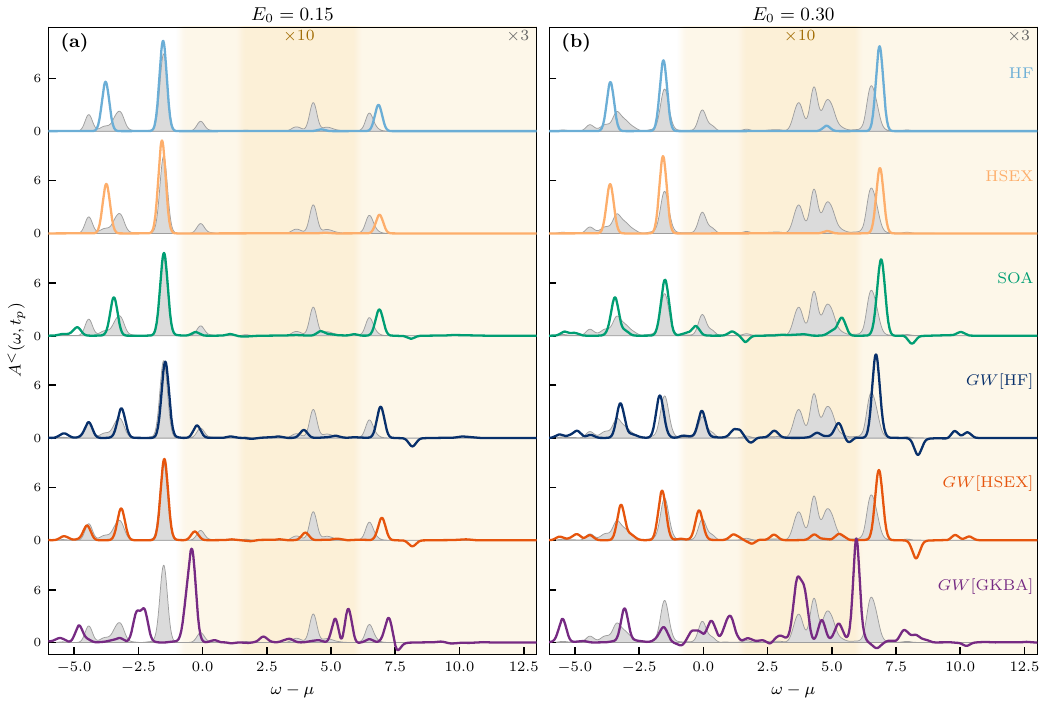}
  \caption{Driven occupied spectral function $A^<(\omega,t_p)$ of the doped four-site chain at $U=2.5$ and $\gamma=0.5$ at probe time $t_p=40$ (pump $\omega_0=8.0$, $t_0=10$, $T_0=2$) for pump amplitudes (a) $E_0=0.15$ and (b) $E_0=0.30$. Gray fill is the exact result. The excited weight above $\mu$ is small compared to the valence weight and is magnified in the shaded regions, with the factors shown at the top of each
  panel. The darker orange region is magnified by a factor of $10$ and the lighter region by a factor of $3$. Spectra displayed at $\delta=5$.}
  \label{fig:driven_spectra_doped}
\end{figure*}
The distinction is structural rather than quantitative. In a static effective potential, removing a particle from an orbital leaves exactly one final state, so that the spectrum consists of one pole per orbital and nothing else. Satellite structures arise because the correlated self-energy $\Sigma^\mathrm{cor}(\omega)$ has poles at the neutral excitation energies of the system, which transfer to the single-particle Green's function as additional poles. A frequency-independent self-energy cannot produce them at any coupling strength, however elaborate its construction. The improvement of the RT-DE over its own reference is thus a direct consequence of the dynamical correlations generated by the second-order scattering term $\Psi$ and the polarization term $\Pi$, and it is not attainable by improving the reference itself.

For the unoccupied spectrum, panel~(b), the ordering changes slightly. SOA is marginally the most accurate over most of the $U$ range, with $GW$[HF] and $GW$[HSEX] being of comparable quality, and only the mean-field approximations are clearly separated from the rest. The advantage of the screened interaction is therefore specific to the occupied spectrum and does not extend to the unoccupied part. A contributing factor is that the SOA contains the second-order exchange diagram, an exchange-type vertex contribution that is absent from $GW$. Such vertex terms partially cancel the self-polarization error, cf.~Sec.~\ref{ss:HSEX_propagators}, and are known to be particularly important for the energies of unoccupied states~\cite{Vlcek2019JCTC, Weng2023JCP}.

$GW$[GKBA] agrees with the other $GW$ variants at weak coupling, but for $U\gtrsim2$ its error increases significantly. At strong coupling it becomes the least accurate method for the occupied spectrum, exceeding even the errors of the mean-field approaches. The corresponding spectrum in Fig.~\ref{fig:gs_spectra} shows strongly broadened and split peaks. The origin is a mismatch between the correlated time-diagonal propagation and the mean-field off-diagonal propagation. The GKBA reference considers a correlated Green's function on the time diagonal, i.e., a non-idempotent density matrix, while the off-diagonal propagation is generated by a mean-field Hamiltonian. A non-idempotent density matrix has fractional diagonal elements in the eigenbasis of the mean-field Hamiltonian and, since its natural orbitals do not coincide with the mean-field eigenstates, also off-diagonal ones, so that the same single-particle energies acquire weight in both the occupied and the unoccupied spectrum, and the weight of each natural orbital is distributed over several eigenvalues. The misplaced weight grows with the interaction strength $U$.

\subsection{Driven doped system} \label{ss:driven_doped}
\subsubsection{Occupied spectral function}

We now turn to the driven dynamics of the doped system. Figure~\ref{fig:driven_spectra_doped} shows the occupied spectral function $A^<(\omega,t_p)$, at probe time $t_p=40$, for the pump amplitudes $E_0=0.15$ and $0.30$. The probe window ($25.9 \le t \le 54.1$ for $\delta=5$) lies entirely after the pump pulse ($t_0=10$, $T_0=2$), while the post-pulse population dynamics discussed in Sec.~\ref{sss:kresolved} is still ongoing. The weight transferred above $\mu$ is small compared to the valence weight and is magnified in the indicated regions so that the differences between the methods become visible.

First, we consider the weaker excitation, $E_0=0.15$. The mean-field approximations resolve only two dominant peaks below $\mu$. They miss the peak located at $\mu$ entirely and do not resolve the structure between $\omega-\mu=-5$ and $-2.5$, which the exact spectrum shows clearly. Above $\mu$ they produce essentially a single dominant peak near $\omega-\mu=6.5$ and strongly underestimate the satellite features in the more strongly magnified window. HF and HSEX differ only by a small shift of the peak positions and by the smaller excited weight of HSEX. At this filling the static screening changes the spectrum only quantitatively but not structurally.

Furthermore, we find that SOA improves upon this. It resolves more than two peaks below $\mu$ and partially reproduces the feature at $\mu$, although the positions of the valence peaks remain somewhat misplaced. The excited part is better described than at mean-field level, but a large part of the structure in the strongly magnified window is still absent.

The mean-field-referenced $GW$ schemes improve significantly further. In particular, the features that SOA begins to resolve are reproduced closer to the exact result. The distinction between the two references carries over from the underlying propagators, and $GW$[HF] as well as $GW$[HSEX] differ in the same way that HF and HSEX do, i.e., by a small shift of the peaks and by the amount of excited weight.

$GW$[GKBA] behaves as in the ground state, Sec.~\ref{ss:eq_benchmark}. The spectrum shows split peaks and misplaced weight, and dominant features that even the mean-field approximations reproduce are found at incorrect positions.

These observations carry over to $E_0=0.30$, where the excited part of the spectrum is significantly more prominent. There, the mean-field approximations show the same deficiencies as before and the mean-field-referenced $GW$ schemes improve with increasing excitation. The latter place the peaks more accurately than at the lower fluence and reproduce the correct number of peaks, although the weight is distributed differently among them, and differently for the HF- and HSEX-referenced variants. Both, together with SOA, produce a peak near $\omega-\mu=10$ that has no clear counterpart in the exact spectrum. The exact spectrum does, however, contain a very weak feature above $\omega-\mu=7.5$, and the peak likely corresponds to this satellite, displaced upward and strongly overestimated.

An additional deficiency of the correlated schemes is that the spectra are not non-negative. While the exact spectrum satisfies the positivity property of Sec.~\ref{ss:spectral_function} and the mean-field references show only minor violations of the order of $10^{-6}$, all correlated schemes carry some negative spectral weight. The violation grows with the driving strength and is localized above the conduction band, in the same spectral region as the satellite structures whose weight is overestimated. The origin is the approximate choice of boundary conditions for the off-diagonal propagation of $G^\gtrless$. The retarded and advanced components $G^\mathrm{R/A}$, and with them the full spectral function $A=A^>+A^<$, are propagated from the exact equal-time anticommutator, cf.~Eqs.~\eqref{eq:GR_init_1} and \eqref{eq:GR_init_2}. Numerically, $A(\omega,t_p)\ge 0$ holds to $10^{-8}$ of the band maximum for every scheme and amplitude considered, whereas $A^<$ alone, which starts from the approximate boundary values \eqref{eq:BC_G} and \eqref{eq:BC_G2} for HF/HSEX or \eqref{eq:BC_G_GKBA} and \eqref{eq:BC_G2_GKBA} for GKBA, reaches $-13\%$ of its maximum ($GW$[HF], $E_0=0.30$). The negative weight amounts to $2$--$4\%$ of the occupied spectral weight at $E_0=0.30$ (SOA $2.2\%$, $GW$[HF] $4.4\%$, $GW$[HSEX] $4.3\%$, $GW$[GKBA] $2.8\%$) and to $0.8$--$2\%$ at $E_0=0.15$. Negative weight in one component is always compensated by the other, i.e., $A^<(\omega,t_p)<0$ implies $A^>(\omega,t_p)>0$.

\subsubsection{Fluence dependence}
\begin{figure}[tb]
  \includegraphics[width=\columnwidth]{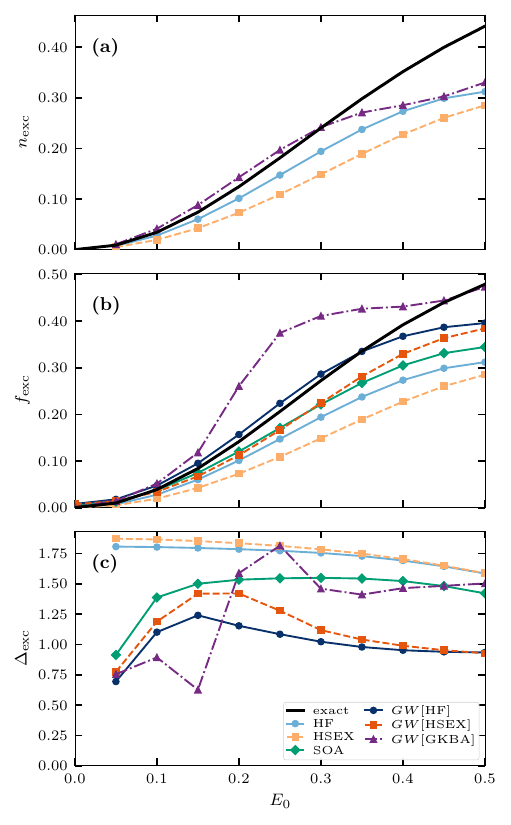}
\caption{%
  Fluence dependence of the driven doped system with four sites at $U=2.5$ and $\gamma=0.5$, for $E_0=0$ to $0.5$. (a) Excited-carrier fraction
  $n_{\mathrm{exc}}=\Delta N_c/N_{\mathrm{el}}$ obtained from the $k$-resolved
  conduction band density. {(b)} Spectral excited fraction
  $f_{\mathrm{exc}}$. {(c)} Conduction-side placement error
  $\Delta_{\mathrm{exc}}$. Scan at $\delta=3$.}
  \label{fig:fluence_scan}
\end{figure}
\begin{figure}[t!]
  \includegraphics[width=\columnwidth]{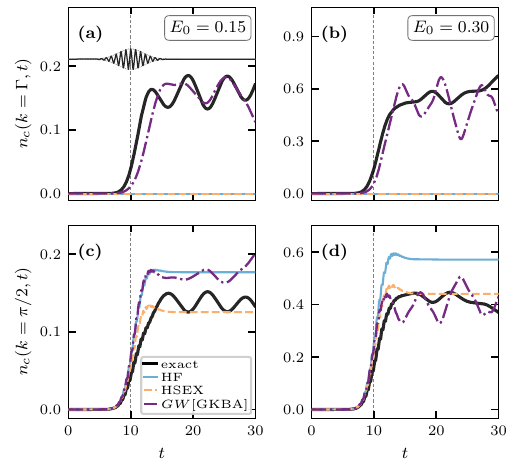}
  \caption{\label{fig:kresolved_density}%
  Momentum- and time-resolved conduction band occupation, $n_c(k,t)$, of the
  doped four-site chain at $U=2.5$ and $\gamma=0.5$ created by the pump ($\omega_0=8.0$, $t_0=10$), at the two
  momenta that carry excited weight: $k=\Gamma$ (top) and $k=\pi/2$ (bottom). Columns correspond to the pump amplitudes $E_0=0.15$ and $0.30$. The vertical line marks the pulse maximum
  ($t_0=10$) and the inset in {(a)} shows the pulse envelope.}
\end{figure}
How the spectral differences between the methods develop with the excitation strength is quantified in Fig.~\ref{fig:fluence_scan}, which shows the three estimators of Sec.~\ref{ss:numerics_observables} over the full amplitude range.

Panel~(a) shows the excited-carrier fraction $n_\mathrm{exc}$ obtained from the time-diagonal Green's function. Only the references appear, since each $GW[\mathrm{X}]$ inherits the value of its reference exactly. The exact system absorbs the most for larger fluences. The GKBA reference follows it most closely for $E_0 \lesssim 0.3$ and then starts to underestimate the absorption. HF lies below it, and HSEX under-excites most strongly. Towards the largest fluences the GKBA result degrades and approaches the HF curve, so that its advantage is largest at weak and intermediate driving.

Panel~(b) shows the excited spectral part  $f_\mathrm{exc}$, which reproduces the ordering already visible in the spectra of Fig.~\ref{fig:driven_spectra_doped}.  $GW$[GKBA] rises far too steeply and overestimates the excited weight substantially throughout the intermediate range, exceeding the exact result before falling back towards it at the strongest fields. $GW$[HF] tracks the exact curve most closely, followed by $GW$[HSEX], which again absorbs too little, then SOA, and finally the two uncorrelated approximations. The correlated reference thus gives the best description of the excited population on the time diagonal and a markedly worse one in the spectrum. Since the two panels derive from the time-diagonal densities and the off-diagonal propagation, their disagreement is a measure of the separation between the two kinds of accuracy.

For the placement error $\Delta_\mathrm{exc}$, panel~(c), we find that its fluence dependence is more structured than that of $n_\mathrm{exc}$ and $f_\mathrm{exc}$. $GW$[HF] is the most accurate over almost the entire range. The $GW$-based schemes first degrade with increasing fluence and then recover at stronger fields. In the intermediate region, where they are least accurate, the difference between $GW$[HF] and $GW$[HSEX] is most pronounced, so that the choice of reference matters most precisely where the schemes are weakest. SOA degrades most strongly in the same region but recovers far less at large fields. HF and HSEX are nearly indistinguishable from one another and carry the largest errors throughout. $GW$[GKBA] starts alongside the other $GW$ schemes and then improves while all others deteriorate, before degrading abruptly to mean-field-level errors, improving once more, and deteriorating again. This nonmonotonic behavior is a consequence of the peak splitting discussed above, combined with the construction of $\Delta_\mathrm{exc}$. Since the measure records displacements of weight relative to $\mu$ in proportion to the displacement, it is highly sensitive to a split feature near the chemical potential. Indeed, the dips and rises of the $GW$[GKBA] curve coincide with the fluences at which a split component of the quasiparticle peak crosses $\mu$ in the corresponding spectra, so that the curve tracks on which side of $\mu$ the artificial splitting appears rather than any systematic change in accuracy.

It is notable that $GW$[GKBA], the scheme with the largest overall spectral error, captures the existence of the excited features---their number and approximate positions---better than the other schemes. Its predictive value is nevertheless limited. Since most of the remaining spectrum is misplaced, the correct and the incorrect parts cannot be separated without an exact reference.

\subsubsection{Momentum-resolved occupations} \label{sss:kresolved}
The momentum-resolved conduction band occupation $n_c(k,t)$, Fig.~\ref{fig:kresolved_density}, identifies where the deficiency of the uncorrelated references is qualitative rather than quantitative.
At $k=\Gamma$ they show no dynamics at all. This follows from momentum conservation as the spatially uniform field couples the two bands at fixed $k$ and cannot populate the initially empty $\Gamma$ point of the conduction band, so that only scattering can do so. Lacking a collision term, the uncorrelated references are insensitive to this momentum entirely, at any driving strength. The GKBA reference follows the exact dynamics closely for weak driving and remains in qualitative agreement under strong driving. At $k=\pi/2$, where interband excitation is directly laser-driven, the static references do show dynamics, but of differing quality. HF over-excites at both fluences while HSEX agrees well with the exact result. Both, however, remain constant after the pulse, again reflecting the absence of scattering. Only the correlated reference reproduces the post-pulse dynamics, although it overestimates the amplitude of the residual oscillations.
\subsection{Driven undoped system} \label{ss:undoped}

\begin{figure*}[t]
  \includegraphics[width=\textwidth]{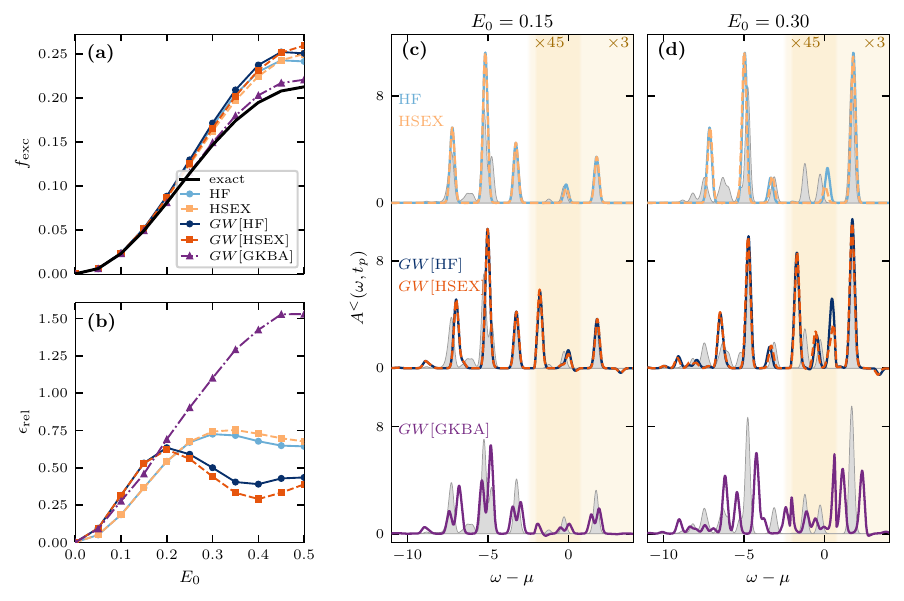}
  \caption{
        Driven benchmark on the undoped four-site chain (interaction $U=2.5$, $\gamma=0.5$ and  pump $\omega_0=5.0$, $t_0=10$,
        $T_0=2$). \textit{Left}: (a) spectral excited fraction $f_\mathrm{exc}$ and (b) relative error $\epsilon_\mathrm{rel}$ of $A^<$, for pump amplitudes $E_0=0$ to $0.5$. \textit{Right}: occupied spectral function $A^<(\omega,t_p)$ at probe time $t_p=40$ for (c) $E_0=0.15$ and (d) $E_0=0.30$, with gray fill giving the exact result. The methods are grouped by reference: HF and HSEX (top), $GW$[HF] and $GW$[HSEX] (middle), and $GW$[GKBA] (bottom). Within each of the upper two rows the two curves are close in the valence region but differ visibly in the magnified window. The weight above $\mu$ is small compared to the valence weight and is magnified in the shaded regions, with the factors annotated at the top of each panel: the darker band is magnified by a factor of $45$, the lighter band by a factor of $3$. Scans at $\delta=3$ and displayed spectra at $\delta=5$.}
  \label{fig:undoped_control}
\end{figure*}

To assess the filling dependence of these results, we now consider the driven undoped system, Fig.~\ref{fig:undoped_control}. As discussed in Sec.~\ref{ss:model}, its ground state is exactly described by the Hartree--Fock solution, so that every deviation appearing at $E_0>0$ is of purely nonequilibrium origin.

The relative error of the occupied spectrum, panel~(b), inverts the ranking of the methods as the pump amplitude is increased. Below $E_0\approx0.2$ the two uncorrelated approximations are the most accurate. Between $E_0=0.2$ and $0.25$ the
curves cross and the mean-field errors saturate and then decline only slowly, whereas the $GW$-based schemes turn over and improve monotonically, reaching their smallest error near $E_0=0.4$, where $GW$[HSEX] is more accurate than $GW$[HF]. The behavior of $GW$[GKBA] is different again. It is the only method whose error grows monotonically over the entire range and reaches values well above those of the mean-field approximations at the largest fluences.

The spectra in panels~(c) and (d) account for the crossover. The magnified window contains a feature near $\omega-\mu=-1.2$ in the exact spectrum, which originates from a final state in which the remaining electrons are left excited (cf.~Sec.~\ref{ss:eq_benchmark}). Neither HF nor HSEX produces it at either fluence, whereas both $GW$-based approaches generate it, displaced to about $-1.7$ and far too strongly. The correlated kernels predict that the feature exists but approximate its position and overestimate its weight. A second feature close to $\mu$ is present in every method. The crossover of $\epsilon_\mathrm{rel}$ follows from the first feature. Between the two amplitudes the exact peak grows by an order of magnitude while the weight placed there by the correlated methods hardly changes. As a pointwise norm, $\epsilon_\mathrm{rel}$ saturates once two features cease to overlap, so that a misplaced and a missing peak lead to comparable penalties while the weight of a feature enters in proportion. Missing the feature altogether is a small error at weak driving and a large one at strong driving, whereas overestimating it costs about the same in both cases. The same reasoning applies to the correlated structure elsewhere in the spectrum.

$GW$[HF] and $GW$[HSEX] place the lower feature at the same energy and with the same weight to better than $1\%$, so that it is generated by the correlated self-energy and is indifferent to which mean-field reference it was built upon.  At the stronger excitation $GW$[GKBA] places this feature closest to the exact position of any method, at about $-1.1$, while producing too little weight rather than too much: position and intensity are described best by different schemes. The feature near $\mu$, by contrast, is placed at too high an energy by every method, and the displacement is not systematically smaller for the more advanced ones.

The overestimation of the lower feature is the expected behavior of a satellite constructed from a single pole of the screened interaction, which places too much weight into the first replica. Thus, the correlated self-energy adds the existence of the feature and not its quantitative description.

The monotonic degradation of $GW$[GKBA] has the same origin as for the ground state. The excitation makes the time-diagonal density matrix increasingly correlated, and the artificial splitting this induces in the off-diagonal propagation grows accordingly. 

It follows from the scan that the two accuracy criteria are not independent but rather anticorrelated, as the property that makes the HSEX reference worse on the time diagonal is precisely what makes the one-shot scheme built upon it better in the spectrum. Above the crossover, $GW$[HSEX] is the most accurate method at every amplitude considered and is systematically better than $GW$[HF], although over exactly the same range the HSEX propagator alone is slightly less accurate than HF.

The mechanism can be attributed to error compensation rather than to a better description. The screened reference transfers less weight into the excited states than the bare-exchange one, and the correlated self-energy builds the features of the excited region from precisely this population while overestimating them. Starting from a smaller population therefore produces a smaller overestimate, and the two errors partially compensate. This interpretation is further supported by the peak positions and weights. The feature near $\mu$, whose weight does depend on the reference, carries visibly less weight in $GW$[HSEX] than in $GW$[HF] and lies closer to the exact value, whereas the feature near $-1.7$, which the kernel generates independently of the reference, is identical in the two schemes.
The advantage is therefore not evidence that HSEX is the better reference since the same under-excitation that improves the spectrum degrades the time-diagonal observables of Sec.~\ref{ss:driven_doped} and the populations of Sec.~\ref{ss:ns40}.

The positivity violation of Sec.~\ref{ss:driven_doped} is present here as well but smaller by a factor of four to six ($0.5\%$ and $0.8\%$ of the occupied weight for $GW$[HF] at $E_0=0.15$ and $0.30$, with a worst value of $-1.7\%$ of the peak), in proportion to the weight of the correlation-induced excited structures, which it accompanies.

Judged by the excited amount alone, panel~(a), the ranking is different once more: all methods slightly over-excite, and $GW$[GKBA] lies closest to the exact result at every amplitude, precisely while its full-spectrum error is the largest of all. The same inversion was found for the doped system in Sec.~\ref{ss:driven_doped}. Neither $\epsilon_\mathrm{rel}$ nor $f_\mathrm{exc}$ alone therefore suffice to assess a scheme, and reference propagators cannot be chosen by benchmarking them against each other: the less accurate of the two uncorrelated approximations yields the more accurate one-shot spectrum.
\subsection{Large systems} \label{ss:ns40}
\begin{figure*}[tb!]
\centering
\includegraphics[width=\textwidth]{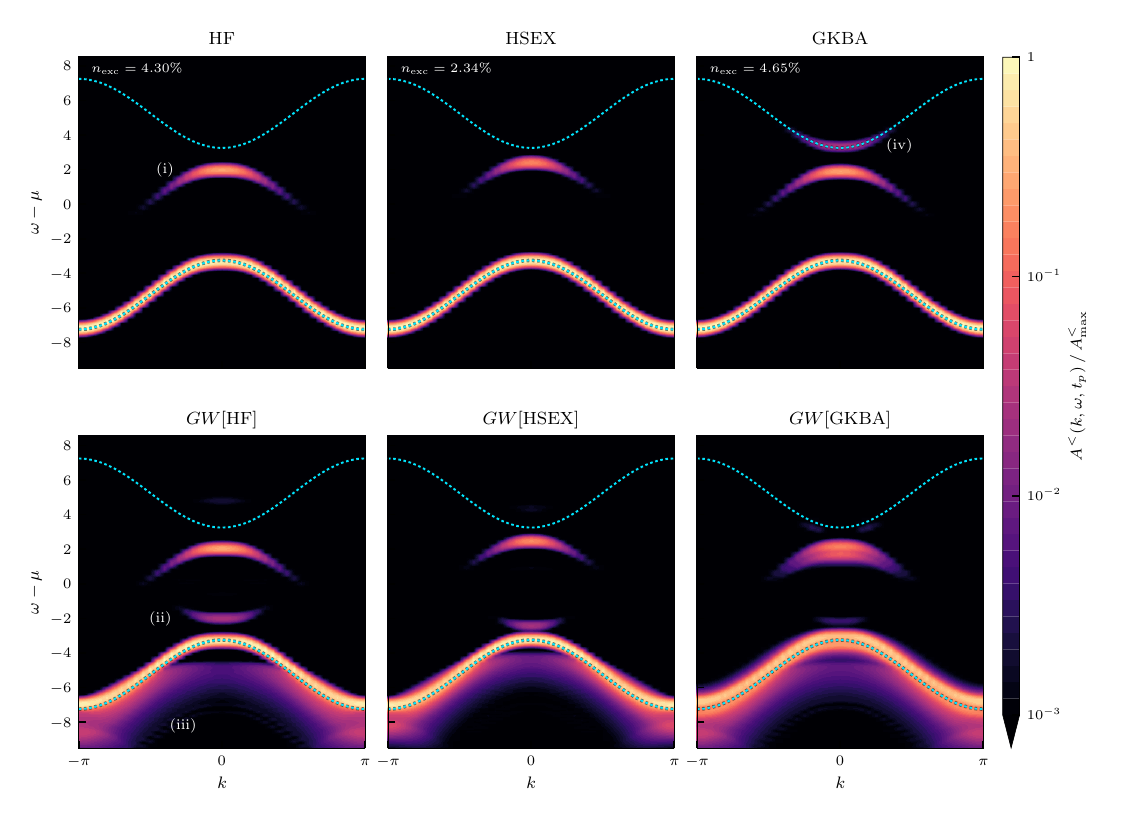}
\caption{%
Momentum-resolved occupied spectral function $A^<(k,\omega)$ of the photoexcited two-band chain with $N_s=40$ sites at $U=2.5$ and $\gamma=0.5$ (logarithmic color scale truncated at $10^{-3}$ of the maximum). Only the positive part is shown due to the logarithmic scale. The excitation is centered at $t_0=10$ and has amplitude $E_0=0.15$ and frequency $\omega_0=5.0$ with width $T_0=2$. The dotted lines show the full ground-state spectrum $A(k,\omega)$ of the system. The spectrum is probed at $t_p=40$ with a probe width of $\delta=5$. The labels (i)--(iv) mark the four nonequilibrium features discussed in the text: (i)~the excitonic replica of the valence band, (ii)~the shake-up satellite, (iii)~the diffuse weight below the valence band, and (iv)~weight at the conduction-band edge of the GKBA reference, attributed to artificial excitations. }
\label{fig:ns40_E0.15contour}
\end{figure*}

\begin{figure*}[tb!]
\centering
\includegraphics[width=\textwidth]{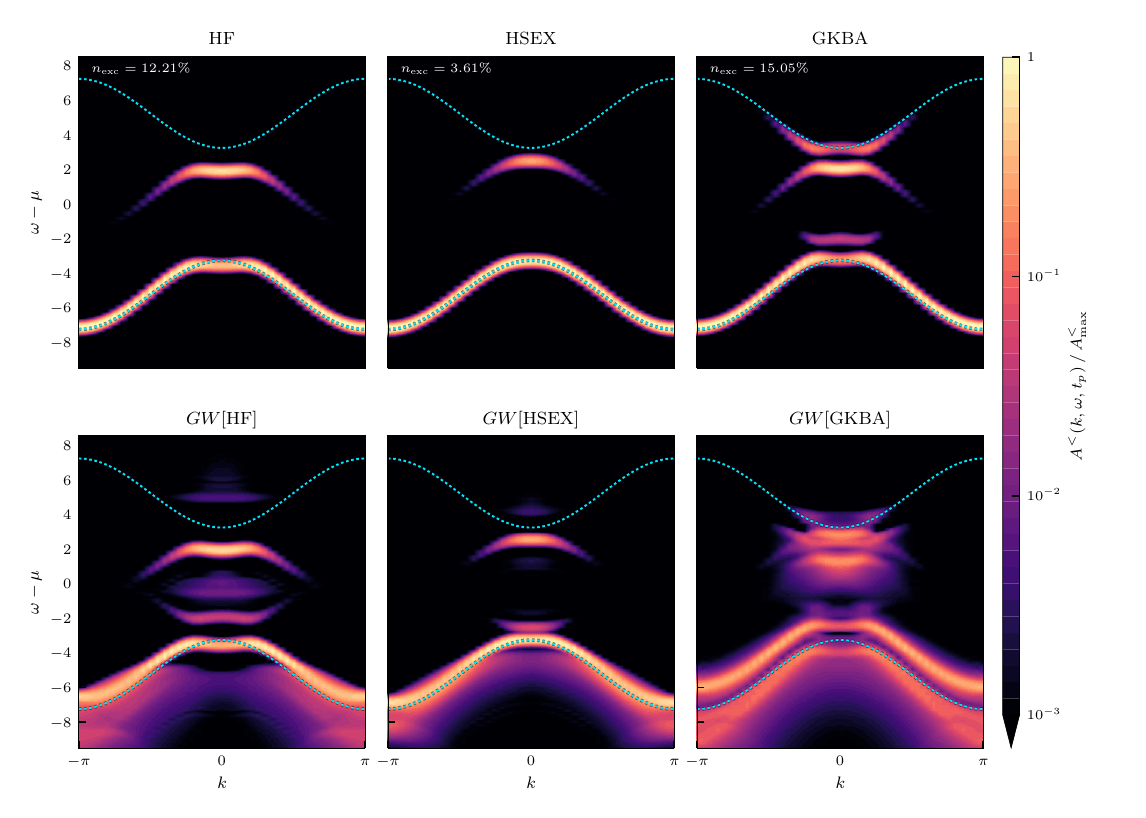}
\caption{%
Same as Fig.~\ref{fig:ns40_E0.15contour}, but at $E_0=0.3$. }
\label{fig:ns40_E0.30contour}
\end{figure*}
In this section we now turn to an undoped system with $N_s = 40$, where the spectra have sufficient momentum resolution. Although exact benchmarks are unavailable at this size, the small-system results show that the correlated kernels reliably predict the existence and position of correlation-induced features, while overestimating their weight. The discussion is therefore restricted to presence, position and qualitative behavior. The parameters are those of the small-system benchmarks ($U=2.5$, $\gamma=0.5$, $t_0=10$, $T_0=2$, $\delta=5$, $t_p=40$), and the system is driven by a sub-gap pulse with $\omega_0=5.0$, i.e., $1.5$ below the quasiparticle gap $E_\mathrm{G}=6.5$ and at the excitonic optical gap $E_\mathrm{G}-E_\mathrm{B}$ (cf.~Sec.~\ref{ss:model}). We again consider the pump amplitudes $E_0=0.15$ and $0.30$. At $E_0=0.30$ the excited fraction, the exciton energy and the band curvature at $\Gamma$ obtained for $N_s=40$ and $N_s=60$ agree ($n_\mathrm{exc}=0.122$ and $E_\mathrm{X}=5.14$ in both cases for HF [not shown]).

Figures~\ref{fig:ns40_E0.15contour} and \ref{fig:ns40_E0.30contour} show the momentum-resolved occupied spectral function $A^<(k,\omega,t_p)$ of the three references (top rows) and of the corresponding RT-DE schemes (bottom rows) at the two amplitudes $E_0$. A logarithmic color scale is used since the structure discussed below carries between one and two percent of the band maximum and is not visible on a linear scale. Due to the logarithmic scale, only the positive part of the spectral function is shown. The correlated schemes do carry negative weight also at this system size. For $GW$[HF] it amounts to $0.4\%$ ($E_0=0.15$) and $1.0\%$ ($E_0=0.30$) of the spectral weight in the displayed window, comparable to the $0.5\%$ and $0.8\%$ of the undoped four-site chain of Sec.~\ref{ss:undoped} and well below the $2-4\%$ of the doped chain of Sec.~\ref{ss:driven_doped}. 

We first consider the spectrum of the system driven at $E_0 = 0.15$, Fig.~\ref{fig:ns40_E0.15contour}. In addition to the valence band, four features are visible [marked in the figure by (i)-(iv)]. (i)~A branch dispersing parallel to the valence band, displaced upward by a constant offset, with its maximum at $(k,\omega-\mu) \approx (\Gamma, 2.0)$. It appears in all six panels and is analyzed in Sec.~\ref{sss:replica}. (ii)~A branch mirroring it below the chemical potential, with its minimum at $(\Gamma, -2.0)$, which appears only for the RT-DE schemes. (iii)~A diffuse weight below the lower valence-band edge, at $\omega-\mu \lesssim -7.3$, likewise present only for the RT-DE schemes. Both features are analyzed in Sec.~\ref{sss:satellite}. (iv)~Weight along the bare conduction band around its minimum at $(\Gamma, 3.3)$, which is present only in the GKBA reference and is attributed to the artificial excitations of the HF-GKBA in Sec.~\ref{sss:replica}.
 
Their different visibility across the panels follows from a single distinction. Every peak of the removal spectrum $\omega(k)$ lies at the difference between the energy of the initial $N$-particle state $E^{(N)}_\mathrm{i}$ and that of the final $(N-1)$-particle state $E^{(N-1)}_\mathrm{f}$,
\begin{align}
  \omega(k) = E^{(N)}_\mathrm{i} - E^{(N-1)}_\mathrm{f}(k)\,, \label{eq:removal_energy}
\end{align}
evaluated at the momentum $k$ of the removed electron, i.e., all features correspond to vertical transitions.  A feature displaced from a band by an excitation energy $E_\mathrm{X}$ can therefore arise in two ways. If the excitation is stored in the initial state, $E^{(N)}_\mathrm{i}= E^{(N)}_0 + E_\mathrm{X}$, every removal energy is shifted upward by the same $k$-independent amount, whereas an excitation left behind in the final state, $E^{(N-1)}_\mathrm{f}(k) = E^{(N-1)}_\mathrm{f,0}(k) + E_\mathrm{X}$, shifts the peak downward as long as the excitation carries $q \approx 0$ and a single energy. In general the two displacements involve different quantities, the excitation energy of the $N$-particle initial state and that of the $(N-1)$-particle final state. That they coincide here, so that a single $E_\mathrm{X}$ can be used for both, is shown in Sec.~\ref{sss:satellite}. The first case requires only an excited population and therefore appears in every scheme, each with its own exciton energy (Sec.~\ref{sss:replica}). The second requires the removal process to end in a correlated final state, which a static self-energy cannot describe. Feature~(i) belongs to the first class, whereas features~(ii) and (iii) belong to the second. Feature~(iv) is not attributed to a removal process of the driven state but to an artifact of the GKBA reconstruction, cf.~Secs.~\ref{ss:HFGKBA} and \ref{sss:replica}.

Schematically, the features (i) and (ii) correspond to the removal processes:
\begin{align}
  \mathrm{(i)}&: 
  |N; \mathrm{X}\rangle \rightarrow |N-1;\, \mathrm{h}_k\rangle,\quad
   \omega(k) = E_v(k) + E_\mathrm{X},
  \label{eq:process_replica}\\
  \mathrm{(ii)}&: 
  |N; \mathrm{e}_k\rangle \rightarrow |N-1; \mathrm{X}\rangle,\quad
  \omega(k) = E_c(k) - E_\mathrm{X},
  \label{eq:process_satellite}
\end{align}
where $|N;\, \mathrm{A}\rangle$ denotes the driven $N$-particle state containing the excitation ``$\mathrm{A}$,'' with ``$\mathrm{X}$'' denoting an exciton with momentum $q \approx 0$, ``$\mathrm{e}_k$'' a photoexcited conduction electron (whose companion hole is understood to be part of the state), and ``$\mathrm{h}_k$'' a valence hole. For the replica the exciton is contained in the initial state and a bare hole remains after the removal of a particle, whereas for the satellite a bare electron is removed and the exciton is left behind. No analogous single line exists for feature~(iii). There the removal of a valence electron leaves the pump-created carriers rearranged within the band, and since these rearrangements form a continuum of momenta and energies rather than a single mode, the displacement in Eq.~\eqref{eq:removal_energy} is not rigid and the branch is smeared into a diffuse tail below the band.

\subsubsection{Excitonic replica} \label{sss:replica}
 
Feature~(i) disperses parallel to the valence band and is displaced upward by a constant offset, which we measure from the equilibrium valence-band maximum, $E_\mathrm{X}\equiv\omega_\mathrm{repl}(\Gamma)-E_v^\mathrm{eq}(\Gamma)$. It is not set by the pump, whose frequency is fixed at $\omega_0 = 5.0$, but depends on the approximation. At $E_0 = 0.15$ it equals $5.21$ for HF, $5.63$ for HSEX and $\sim5.1$ for GKBA. It is therefore an internal energy of the system, namely that of the neutral electron--hole excitation created by the pump. At finite density it also contains the pump-induced renormalization of the band from which the replica derives. The offset measured from the driven valence band, denoted $\Delta_\mathrm{X}$ below, separates this out (Sec.~\ref{sss:fluence}). With the quasiparticle gap $E_\mathrm{G} = 6.5$ of the undoped system, cf.~Sec.~\ref{ss:model}, these offsets correspond to exciton binding energies $E_\mathrm{B} = E_\mathrm{G} - E_\mathrm{X}$ of $1.29$ (HF), $0.87$ (HSEX) and $\sim1.4$ (GKBA). The RT-DE schemes follow their respective references, with offsets larger by $0.05$ ($GW$[HF]: $5.26$, $GW$[HSEX]: $5.68$), so that the exciton energy is set by the reference propagator and only weakly influenced by the $GW$ self-energy. Here $E_\mathrm{B}$ is defined relative to the exact quasiparticle gap, which all references share in equilibrium (Sec.~\ref{ss:model}), so that the differences between the references are differences in the exciton energy alone. The excitons are strongly bound, with $E_\mathrm{B}$ amounting to $13-22\%$ of the gap. This ratio is comparable to that of excitons in monolayer transition-metal dichalcogenides~\cite{chernikov_prl_14}. The smaller binding energy of the HSEX reference is a first indication of the photocarrier screening analyzed in Sec.~\ref{sss:fluence}.

These observations identify feature~(i) as the excitonic replica of the valence band, familiar from the theory and experimental observations of excitonic sidebands in time-resolved photoemission of photoexcited semiconductors~\cite{Perfetto2016,StefanucciPerfetto2021,StefanucciPerfetto2026,man2021}. It arises through process~(i) of Eq.~\eqref{eq:process_replica}. Removing the conduction electron of a $q=0$ exciton leaves the same one-hole final state as valence removal from the ground state, while the initial state lies higher by $E_\mathrm{X}$. The branch therefore copies the valence-band dispersion, its momentum dependence entering only through the final-state hole, and its intensity is set by the excited population, $f_\mathrm{exc}\approx n_\mathrm{exc}$ for the HF and HSEX references at both amplitudes.
 
For the GKBA reference, however, $\sim10\%$ ($E_0=0.15$) to $\sim20\%$ ($E_0=0.30$) of its excited spectral weight lies not on the excitonic branch but at the bare conduction-band edge. We again attribute this weight to the artificial excitations of the HF-GKBA discussed in Sec.~\ref{ss:HFGKBA}. It appears at the eigenvalue of $h^\mathrm{HF}[G^\mathrm{GKBA}]$, where the mean-field reconstruction places any incoherent population regardless of its binding, whereas an incoherent exciton population contributes at the replica energy in the exact spectrum~\cite{StefanucciPerfetto2021}, and free carriers at the band edge cannot be generated directly by the sub-gap pump. Correspondingly, the feature is largely suppressed in $GW$[GKBA], where the off-diagonal propagation is improved, and contributes to the broadening instead. A physical free-carrier population created by exciton--exciton scattering at these densities cannot be excluded at the present level, but would not be represented differently by the mean-field reconstruction.
 
Since only an excited population and the electron--hole attraction of the reference are required, the replica is present already at the mean-field level \cite{Perfetto2019PRM, Perfetto2020PRB}. 
\subsubsection{Shake-up satellites} \label{sss:satellite}
 
The branch below $\mu$ appears only for the correlated schemes, and its origin is the dynamical part of the self-energy. In the interacting driven system, following the removal of an electron, the remaining $N-1$ electrons can be left excited, and the photoelectron then carries correspondingly less energy. Here the removal of a photoexcited conduction electron with momentum $k$ leaves the remaining $N-1$ electrons with an additional exciton of momentum $q \approx 0$, created out of the still nearly filled valence band, which places the satellite, feature~(ii), at $E_c(k) - E_\mathrm{X}$, cf.~Eq.~\eqref{eq:process_satellite}.
 
In principle, any low-lying neutral excitation can be shaken off in this way, and for a long-range interaction plasmon-type charge fluctuations are the natural alternative candidate. In the undoped ground state of the present model, however, the RPA polarizability vanishes identically, since the density is band-diagonal and the valence band is filled, so that $W=w$ and the reference carries no collective charge mode. The charge fluctuations of the driven system are those of the photocarriers. The following observation identifies the shaken-off mode as the $q\approx0$ exciton. The offset is the exciton energy $E_\mathrm{X}$ and not the pump frequency. Measured from the equilibrium conduction-band minimum, the satellite of $GW$[HF] at $E_0 = 0.15$ lies $5.27$ below it, matching the replica offset $E_\mathrm{X}=5.26$, whereas the pump frequency is $5.0$. The two processes are distinct, and their offsets need not necessarily coincide. The replica offset is the energy of the exciton stored in the initial $N$-particle state, the satellite offset that of the exciton created in the final $(N-1)$-particle state. Both excitons, however, form in the same photocarrier background, and the two states differ only by the removed electron, an $\mathcal{O}(1/N_s)$ perturbation, so that the offsets are expected to agree and, in particular, to shift together with the excited density. This is what is observable as they agree to better than $0.01$ at both amplitudes and both decrease by $0.10$ between $E_0=0.15$ and $0.30$. In the considered particle--hole-symmetric system the two branches moreover mirror each other about $\mu$ by band symmetry.  The shaken-off mode is thus the same $q \approx 0$ exciton identified in Sec.~\ref{sss:replica}, and the satellite is exciton shake-up rather than a replica generated by the pump photon. The rigidity of the offset also determines the momentum of the excitation left behind as removing an electron of momentum $k$ while creating an excitation of momentum $q$ leaves the photoelectron at $k-q$, so a replica at the same $k$ as its parent requires $q \approx 0$. A dispersing excitation, such as a plasmon branch, would smear the branch rather than copy the band. Further, the satellite weight is driven by the excited population. Consequently, it grows significantly when increasing the amplitude from $0.15$, in Fig.~\ref{fig:ns40_E0.15contour}, to $0.3$, in Fig.~\ref{fig:ns40_E0.30contour}. The weight is practically absent at weaker excitation (not shown), the same population-driven behavior as the high-energy peak identified in the excited doped system, cf.~Sec.~\ref{ss:driven_doped}. 
 
The weight below the valence band arises analogously. A valence band electron is removed, and the remaining electrons are left excited. Because the interaction conserves the two band occupations separately, cf.~Sec.~\ref{ss:model}, the excitation that is left behind cannot be an interband one. It is, instead, an intraband rearrangement of the pump-created carriers, and such rearrangements form a dense set of final states covering a range of energies rather than a single mode. The result is a diffuse tail below the band instead of a sharp copy of it.
 
The small-system benchmark of Sec.~\ref{ss:undoped} suggests that weight in these spectral regions is physical---the exact spectrum carries weight in the same regions that the mean-field approximations leave empty---and also that the one-shot schemes overestimate its magnitude considerably. This behavior is expected to persist here, and we therefore restrict the discussion of the final-state features to their presence and position.
  
\subsubsection{Fluence and reference-propagator dependence}
\label{sss:fluence}
 
At $E_0 = 0.30$, Fig.~\ref{fig:ns40_E0.30contour}, the same features persist and the differences between the references become more pronounced. The first difference is the absorbed density itself. Under identical driving the references reach $n_\mathrm{exc} = 12.2\%$ (HF), $3.6\%$ (HSEX) and $15.1\%$ (GKBA), with each RT-DE scheme inheriting the value of its reference. The HSEX absorption saturates as the photoexcited carriers screen the exchange, and the weakened electron--hole attraction reduces the absorption. This leads to a saturation of the excited fraction near $3.6\%$, regardless of further increase of the drive. Part of the reduced HSEX absorption is detuning. The pump at $\omega_0=5.0$ lies $0.6$ below the HSEX exciton, but only $0.2$ ($0.1$) below the HF (GKBA) exciton. Detuning alone, however, does not lead to the observed saturation. The saturation reflects the blue shift of the exciton with density ($E_\mathrm{X}$ grows from $5.63$ to $5.73$ for the HSEX reference and from $5.68$ to $5.79$ in $GW$[HSEX]), which increases the detuning as carriers accumulate---the screening feedback itself. One consequence has to be highlighted. Comparisons between the HF and HSEX families, at fixed $E_0$, conflate the choice of reference with a more than threefold difference in excitation density and are therefore avoided in the following. 
 
Another difference is the direction of the shift of the excitonic replica~(i) with increasing drive. Between $E_0=0.15$ and $0.30$ the HF replica moves down by $0.07$, whereas the HSEX replica moves up by $0.10$ (by $0.38$ between $E_0=0.05$ and $0.30$). The two references predict opposite signs for the same observable under the same drive. To interpret this, we write the replica position as $\omega_\mathrm{repl}(k)=E_v(k)+\Delta_\mathrm{X}$, where $E_v(k)$ is the driven valence band extracted from the same spectrum
(Sec.~\ref{sss:mass}) and $\Delta_\mathrm{X}$ is the offset from it, so that $E_\mathrm{X}=\Delta_\mathrm{X}+E_v(\Gamma)-E_v^\mathrm{eq}(\Gamma)$. $\Delta_\mathrm{X}$ is a property of the reference alone as it agrees between HF and $GW$[HF], and between HSEX and $GW$[HSEX], to better than $0.01$ at every amplitude, while the kernel shifts the driven band and the replica together by a few hundredths. For HF the two contributions shift in opposite directions. The driven valence band moves down by $0.22$ (band-gap renormalization), while $\Delta_\mathrm{X}$ grows from $5.30$ to $5.45$, so that $E_\mathrm{X}$
decreases only from $5.21$ to $5.14$ and the small red shift is the residual of a sinking band and an increasing excitation energy. For HSEX the driven valence band effectively does not change ($\lesssim 0.01$), while $\Delta_\mathrm{X}$ grows from $5.66$ to $5.77$, so that the entire blue shift of the replica reflects the weakening of the electron--hole attraction by photocarrier screening.
 
At the larger amplitude the replica dispersion itself is visibly reshaped. There the branch top in the $GW$[HF] panel of Fig.~\ref{fig:ns40_E0.30contour} is flattened for $|k| \lesssim 0.2\pi$ relative to the low-fluence dispersion of Fig.~\ref{fig:ns40_E0.15contour}. This reshaping  will be discussed in more detail in the following section.
 
Finally, the GKBA-referenced scheme degrades qualitatively between the two amplitudes. At $E_0 = 0.15$ its excited feature is a broadened but recognizable branch. However, at $E_0 = 0.30$ no single dispersion can be assigned, and its dominant excited weight no longer coincides with any feature of its own reference. This is the momentum-resolved counterpart of the monotonic error growth of $GW$[GKBA] with fluence found in the small-system benchmark, Fig.~\ref{fig:undoped_control}(b), and has the same origin. Summed over momentum, the individual splittings merge into a broadening of the valence band, as already seen at small $N_s$ in Fig.~\ref{fig:undoped_control}. The two static references show no such effect. Their RT-DE approaches leave the coherent valence structure essentially unchanged and add weight only in the gap and below the valence band, where the references themselves carry an order of magnitude less.
 
\subsubsection{Reshaping of the driven bands}
\label{sss:mass}

\begin{figure}[tb!]
\centering
\includegraphics[width=\columnwidth]{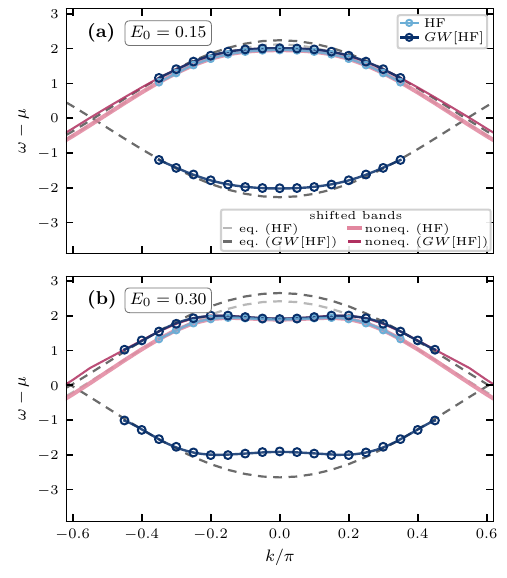}
    \caption{Dispersions of the excitonic replica, feature~(i), and the shake-up satellite, feature~(ii), at $N_s = 40$. At each momentum, the symbols mark the peak position of $A^<(k,\omega,t_p)$ along $\omega$, for $GW$[HF] (dark) and for the HF replica (light), for (a) $E_0 = 0.15$ and (b) $E_0 = 0.30$. Dashed: the dilute-limit constructions
    $E_v^\mathrm{eq}(k) + E_\mathrm{X}$ and $E_c^\mathrm{eq}(k) - E_\mathrm{X}$ from the equilibrium bands, shifted rigidly to match
    each branch in the wings, $|k| \ge 0.30\pi$. Solid: the same constructions built from the driven bands extracted from the same spectrum, $E_v(k)+\Delta_\mathrm{X}$ and $E_c(k)-\Delta_\mathrm{X}$, shifted to the measured $\Gamma$ position of the branch. Both are shown for HF and for $GW$[HF].}
\label{fig:ns40_dispersion}
\end{figure}

The flattening noted above is the lattice-model counterpart of an effect established for photoexcited semiconductors. With growing exciton density the excitonic sideband and the band it derives from are reshaped from a dome into a Mexican-hat form. The mechanism has been derived analytically for a statically screened mean field and evaluated for monolayer materials \cite{StefanucciPerfetto2026}, and the reshaped dispersion has been resolved in time- and angle-resolved photoemission of a monolayer semiconductor, where it appears both in the valence band and in its replica \cite{Pareek2026}. Related Mexican-hat-type dispersions are a known equilibrium fingerprint of excitonic condensation \cite{Wakisaka2009,Kaneko2013}. The pump creates coherent electron--hole pairs, and the resulting anomalous (pairing-type) contribution to the mean field reshapes the bands in the same way as the order parameter of an excitonic insulator, which is why the effect appears already at the Hartree--Fock level. Correlations beyond mean field enhance the curvature change but do not create it, consistent with earlier observations within the RT-DE~\cite{reeves2025} and with two-time Kadanoff--Baym studies of photoexcited excitonic insulators~\cite{Golez2016}.

Figure~\ref{fig:ns40_dispersion} shows the dispersions of features~(i) and (ii), extracted as the peak position of $A^<(k,\omega,t_p)$ at each momentum, together with two constructions. The dashed curves are the equilibrium bands displaced by the exciton energy, $E_v^\mathrm{eq}(k) + E_\mathrm{X}$ and $E_c^\mathrm{eq}(k) - E_\mathrm{X}$, i.e., the dilute-limit replicas. Further, the solid curves are the corresponding constructions built from the \emph{driven} bands extracted from the same spectrum. Comparing the branches with each other, in turn, separates two effects that are easily conflated.
 
Against the driven bands, the branches do not reshape at all. Shifting the extracted driven valence band rigidly to meet the sideband at $k =\Gamma$, the difference is at most $3\times 10^{-3}$ for all considered $k$. The sideband is thus a rigid replica, in the strict sense, of a parent band that has itself been reshaped by the excitation. The rigidity follows because the valence peak and its replica share the same one-hole final states and differ only by the $k$-independent initial-state energy $E_\mathrm{X}$. 
 
Against the equilibrium bands the branches do depart, due to the level repulsion between the exciton replica of the valence band, $E_v(k)+E_\mathrm{X}$, and the conduction band $E_c(k)$, whose detuning near $\Gamma$ is $E_c-E_v-E_\mathrm{X}=E_\mathrm{B}\approx1.3$ and whose coupling is proportional to the exciton amplitude at $k$~\cite{StefanucciPerfetto2026,Pareek2026}. To make it observable, the dashed curves are anchored not at $\Gamma$ but for $|k| \ge 0.30\pi$, outside the momentum range over which the exciton weight is appreciable. Approaching $\Gamma$, each branch departs from its equilibrium construction in the direction of the chemical potential. The sideband, which lies below the conduction band, is pushed down by $0.16$ ($E_0=0.15$) and $0.53$ ($0.30$) for HF. Further, the satellite, which lies above the valence band, is pushed up by a similar amount. The effect is present already at the mean-field level and is about $1.4$ times larger in $GW$[HF], and its magnitude follows the exciton weight, so that it peaks at $\Gamma$ and vanishes in the wings by construction of the anchor. 

\begin{figure}[tb!]
\centering
\includegraphics[width=\columnwidth]{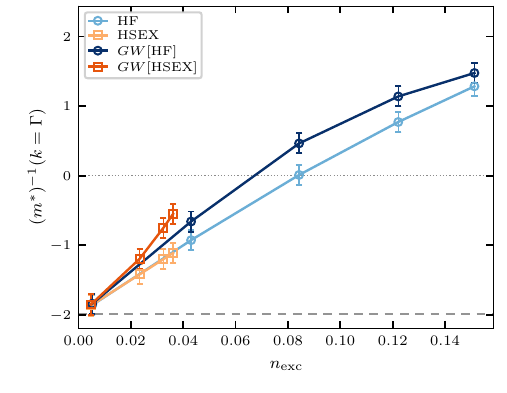}
\caption{Inverse effective mass at the valence band center,
 $(m^*)^{-1}(k=\Gamma) = \partial^2\omega/\partial k^2|_{k=\Gamma}$ in units of
 $Ja^2/\hbar^2$, as a function of the excited fraction
 $n_\mathrm{exc}$ at the probe time, for pump amplitudes $E_0 = 0.05$
 to $0.45$ ($N_s = 40$, $\omega_0 = 5.0$, $t_p = 40$, $\delta = 5$).
 Each RT-DE scheme is plotted at the excited fraction of its
 reference. The dashed line is the value $-2.0$ of the ground-state band. Error bars give the difference between the  five-point and seven-point estimates of the curvature, i.e., the  sensitivity to the fit window. The GKBA reference and $GW$[GKBA]  are omitted.}
\label{fig:ns40_mass}
\end{figure}
 
We characterize the reshaping quantitatively by the inverse effective mass at the band center, 
\begin{align}
    (m^*)^{-1}(k=\Gamma) = \frac{\partial^2\omega}{\partial
k^2}\Bigg|_{k=\Gamma},
\end{align}
obtained from the extracted dispersion as follows. A branch is tracked outward from $k=\Gamma$ by localizing the maximum within a fixed interval around the position found at the preceding momentum, so that only continuity of the branch is assumed. Each maximum is then refined between grid points by a parabola through
its three neighboring samples, which is necessary because the finite probe window band-limits $A^<$ and the discrete $\omega$ grid would otherwise discretize the curvature. The inverse mass is the least-squares parabola through $k/\pi = 0, \pm 0.05, \pm 0.10$, and the error bars in Fig.~\ref{fig:ns40_mass} are the difference to the corresponding seven-point estimate, i.e., they measure the sensitivity
to the fit window. The GKBA reference and $GW$[GKBA] are excluded from this analysis as their excited feature is the split multiplet discussed above, for which a single curvature at $\Gamma$ is not defined.
 
Figure~\ref{fig:ns40_mass} shows the curvature as a function of the excited fraction. In the dilute limit the branches are rigid replicas of the equilibrium bands as well. A rigid copy of the equilibrium valence band carries that band's own inverse mass, $-2.0$ for the dispersion used here, and extrapolating the four curves linearly to $n_\mathrm{exc}\to 0$ gives $-1.99$ (HF), $-1.98$ (HSEX), $-2.00$
($GW$[HF]) and $-2.02$ ($GW$[HSEX]) (in units of $Ja^2/\hbar^2$). It should be noted that the softening sets in immediately as already at $n_\mathrm{exc} \approx 0.005$ the measured values lie about $0.13$ above $-2.0$, so the equilibrium rigid replica is a limit that the trend approaches rather than a regime the calculations occupy. At $E_0 = 0.15$, the dispersion has flattened to $-0.66$, i.e., by a factor of three.
 
The dispersion softens monotonically with the excited density and inverts sign. For the HF-based approaches the zero crossing lies at $n_\mathrm{exc} \approx 8.4\%$ for the reference and $6.7\%$ for RT-DE. Beyond these densities the branch top at $\Gamma$ is a local minimum and the dispersion has acquired the Mexican-hat form of Fig.~\ref{fig:ns40_dispersion}(b), with maxima near $|k| \approx
0.2\pi$.
 
The inversion is a property of the reference. It occurs in the bare HF reference, which contains no correlation kernel, produces no satellite, and has no screened interaction at any level. The one-shot approaches inherit the trajectory and shift it slightly, $GW$[HF] lying about one error bar above HF at every shared density and crossing zero slightly earlier. The reshaping is thus determined by the driven mean-field bands, in the same way as the existence, offset and intensity of the replica itself, and it is controlled by the excitation density rather than by the treatment of exchange. The HSEX family, which never reaches the inversion at the amplitudes considered (Sec.~\ref{sss:fluence}), follows the same trajectory where the densities overlap. Comparing at matched density, HSEX at $n_\mathrm{exc} \approx 3.6\%$ gives $-1.11$ against $-0.93$ for HF at $4.3\%$, and the corresponding kernels give $-0.55$ against $-0.66$. The two families are consistent within the error bars. 
 
There are, however, some limitations. The density at which the inversion occurs is not a transferable number. Varying the range of the interaction shifts it in proportion to the exciton binding energy, and breaking the symmetry between the bands moves it independently of that, so the threshold depends on at least two model parameters and should be read as a property of this system. Moreover, the analysis is restricted to the HF and HSEX families for the reason given above. 
\section{Conclusion} \label{s:conclusion}

In this work, we formulated and implemented the real-time Dyson expansion (RT-DE) for the nonequilibrium $GW$ self-energy using reference propagators of general form with time-local off-diagonal evolution. We considered three choices: Hartree--Fock (HF), Hartree plus statically screened exchange (HSEX) with screening updated during the nonequilibrium propagation, and a reference based on the HF-GKBA. The resulting formulation retains the linear scaling with propagation time of the original RT-DE while incorporating dynamical screening into the nonequilibrium spectra.

Benchmarks against exact diagonalization show that the HF- and HSEX-referenced variants remain reliable at the $GW$ level, for the small systems considered, both in equilibrium and after excitation. In each case, the RT-DE reduces the spectral error of the underlying reference over the investigated ranges of interaction strength and pump amplitude. It also recovers satellite structures that are absent from the static reference spectra. Their weights are nevertheless significantly overestimated, which is consistent with a known limitation of the $GW$ self-energy~\cite{golze_gw_2019}. The correlated kernel, therefore, captures the occurrence and approximate energies of these satellites more reliably than their spectral weights.

The GKBA-based reference behaves qualitatively differently. It describes scattering-induced carrier dynamics that the HF and HSEX references cannot describe in the homogeneous system, but yields the least accurate spectra among the schemes tested. This behavior is due to the mean-field off-diagonal reconstruction of the HF-GKBA rather than to its correlated time diagonal. Representing a correlated state in the eigenbasis of the mean-field Hamiltonian assigns weight at the same single-particle energies to the occupied and unoccupied spectra. Within the RT-DE, these artificial excitations appear as broadened and split peaks. On the other hand, HSEX underestimates the excited population but gives the most accurate spectra for the driven undoped system above the crossover discussed in Sec.~\ref{ss:undoped}. In that regime, its population error partially compensates the overestimated satellite weight generated by $GW$. The accuracy of an RT-DE calculation is thus determined by the combination of reference and self-energy, not by the reference alone. In particular, accuracy on the time diagonal does not imply accuracy in the spectrum.

The calculations for a photoexcited chain with $N_s=40$ demonstrate that RT-DE can resolve correlation-induced spectral features at sizes beyond the reach of exact diagonalization. These features can be distinguished by whether the excitation energy is carried by the initial or the final state of the photoemission process. The excitonic replica of the valence band belongs to the first category and is already present in the reference spectrum. Its offset, intensity, and dispersion are largely fixed by the reference and inherited by the corresponding $GW$ calculation. This includes the density-induced evolution of the driven bands from a dome-shaped to a Mexican-hat dispersion~\cite{StefanucciPerfetto2026,Pareek2026}. The associated curvature inversion at $\Gamma$ occurs near $n_\mathrm{exc}\approx7-8\%$, with only small differences between each reference and its RT-DE counterpart. By contrast, the shake-up satellite and the diffuse weight below the valence band involve excited final states and appear only after inclusion of the correlated kernel. Such final-state features distinguish a correlated spectrum from a mean-field one. With time-resolved photoemission now resolving the excitonic replica, its density dependence and its hybridization with the bands~\cite{man2021,dong2021,Pareek2026}, the accompanying satellites are within experimental reach, and their identification requires a correlated description of the removal process~\cite{StefanucciPerfetto2021,StefanucciPerfetto2026}.

Several extensions of the reference propagator are possible. For example, one could retain only the nonequilibrium change of a statically screened self-energy~\cite{attaccalite2011,attaccalite2013,perfetto2020}, or combine a separately calculated quasiparticle band structure with a time-local COHSEX self-energy~\cite{hedin_new_1965}, as in time-dependent adiabatic $GW$ approaches~\cite{ChanQiu2021,ChanQiu2023,Hou2025,Perfetto2015NEQBSE,Sangalli2021}. A more consequential improvement would be a correlated time-local reference that describes scattering without introducing the artificial spectral excitations of the HF-GKBA.

The same benchmarks show the limitations of time-local (adiabatic) $GW$ approaches~\cite{ChanQiu2021,ChanQiu2023,Hou2025} with regard to what they can deliver for spectra. Their self-energy is frequency independent, so the spectral function contains one pole per orbital, however the static screening is constructed (Sec.~\ref{ss:eq_benchmark}). Consequently, the excitonic replica and the band reshaping are reproduced, as by HF and HSEX here, but satellites and shake-up features are absent by construction. Adiabatic $GW$ therefore effectively yields mean-field spectra with renormalized bands.

The derivation in Sec.~\ref{ss:RTDE_GW} is not restricted to the $GW$ self-energy. The same construction can be applied to particle--particle and particle--hole $T$-matrices and to the dynamically screened ladder approximation~\cite{joost_prb_22}. SOSEX-type vertex corrections could reduce both the excessive satellite weight and the self-polarization error discussed in Sec.~\ref{ss:HSEX_propagators}, but their dynamical character would reintroduce the memory dependence avoided by the time-local formulation. The positivity violations found in Secs.~\ref{ss:driven_doped} and \ref{ss:undoped} have a different origin. Although positive-semidefinite diagrammatic constructions are available~\cite{Stefanucci2014PSD,Uimonen2015PSD,Pavlyukh2016PSD}, positivity in the present RT-DE formulation is determined by the approximate boundary conditions rather than solely by the choice of self-energy diagrams. For the undoped chain the violations are of comparable size for the small and the large system (at most $1\%$ of the occupied weight for $GW$[HF]), and a factor of four to six smaller than in the doped four-site chain. For the cases studied they primarily affect spectral weight rather than the location and overall structure of the spectral features.

The primary obstacle to applications beyond model systems is the scaling with the basis size $N_b$. Storing the four-index function $\mathcal{G}_{ijkl}$ requires $\mathcal{O}(N_b^4)$ memory, while the contractions in Eq.~\eqref{eq:G2_GW_MF_propagators_1} scale as $\mathcal{O}(N_tN_b^5)$ for the density--density interactions used here and as $\mathcal{O}(N_tN_b^6)$ for a general interaction tensor. Even after exploiting symmetries such as translational invariance, this limits direct calculations to relatively small basis sizes. A promising route toward material-specific applications is to combine RT-DE with a low-rank representation of the two-particle function. In the $\delta$NEGF approach~\cite{schroedter_26}, for example, the four-index function is replaced by an ensemble of single-particle fluctuations with memory scaling $\mathcal{O}(N_rN_b^2)$, where $N_r$ denotes the approximate rank. Stochastic sampling has already enabled time-diagonal calculations with basis sizes of order $10^4$~\cite{schroedter_26}. Combining this approach with the RT-DE could extend $GW$-level nonequilibrium spectroscopy to realistic basis sizes. This would permit direct comparisons with time- and angle-resolved photoemission measurements of excitonic replicas, shake-up satellites, and density-dependent band reshaping in photoexcited semiconductors~\cite{man2021,dong2021,Pareek2026}.

\begin{acknowledgments}
We acknowledge valuable comments by Christopher Makait. This work has been supported by the Deutsche Forschungsgemeinschaft (DFG, German Research Foundation) via project no. 464370560. This material is based upon work supported by the U.S. Department of Energy, Office of Science, Office of Basic Energy Sciences, Computational and Theoretical Chemistry program under Award Number DE-SC0026045. 
\end{acknowledgments}
\bibliography{library,mb-ref-2}
\end{document}